\documentclass[12pt]{article}
\usepackage{amsmath,amssymb}
\usepackage[pdftex]{graphicx}
\usepackage{psfrag,epsf}
\usepackage{enumerate}
\usepackage{subcaption}
\usepackage{wrapfig}
\usepackage{indentfirst}
\usepackage{url}
\usepackage[utf8]{inputenc}
\usepackage[table]{xcolor}
\usepackage[colorlinks,citecolor=blue,urlcolor=blue]{hyperref}
\usepackage{natbib}
\usepackage{url} 
\usepackage{threeparttable}
\allowdisplaybreaks
\usepackage[ruled,linesnumbered]{algorithm2e}
\usepackage{algorithmic}
\usepackage{newtxmath}
\newcommand{\blind}{1}

\DeclareMathOperator{\tr}{tr}
\begin{document}

\bibliographystyle{JASA}
\setlength{\bibsep}{0.25pt} 
\renewcommand{\bibfont}{\footnotesize} %

\def\spacingset#1{\renewcommand{\baselinestretch}%
{#1}\small\normalsize} \spacingset{1}


\if1\blind
{
  \title{\bf Joint Latent Space Model for Social Networks with Multivariate Attributes}
  \author{Selena Shuo Wang\thanks{
     \textit{This research is partially supported by a grant from National Science Foundation under Grant No. DMS 1830547. The authors would like to thank Prof. Vishesh Karwa of Temple University, Prof. Srijan Sengupta of North Carolina State University and Prof. Jessica Logan of The Ohio State University for discussions that helped in conceptualizing the statistical models. 
}}\hspace{.2cm},
    Department of Psychology\\
    Subhadeep Paul, 
    Department of Statistics \\ 
        Paul De Boeck,
    Department of Psychology \\\\
    The Ohio State University}
    \date{}
  \maketitle
} \fi

\if0\blind
{
  \bigskip
  \bigskip
  \bigskip
  \begin{center}
    {\LARGE\bf Joint Latent Space Model for Social Networks with Multivariate Attributes}
\end{center}
  \medskip
} \fi

\bigskip

\begin{abstract}

In many application problems in social, behavioral, and economic sciences, researchers often have data on a social network among a group of individuals along with high dimensional multivariate measurements for each individual. To analyze such networked data structures, we propose a joint Attribute and Person Latent Space Model (APLSM) that summarizes information from the social network and the multiple attribute measurements in a person-attribute joint latent space. We develop a Variational Bayesian Expectation-Maximization estimation algorithm to estimate the posterior distribution of the attribute and person locations in the joint latent space. This methodology allows for effective integration, informative visualization, and prediction of social networks and high dimensional attribute measurements. Using APLSM, we explore the inner workings of the French financial elites based on their social networks and their career, political views, and social status. We observe a division in the social circles of the French elites in accordance with the differences in their individual characteristics.

\end{abstract}

\vspace{ 6mm}

\noindent%

\spacingset{1.25} 
\section{Introduction}

Understanding interactions among sets of entities often represented as complex networks, is a central research task in many data-intensive scientific fields, including Statistics, Machine learning, Social sciences, Biology, Psychology, and Economics \citep{watts1998collective,barabasi1999emergence,albert2002statistical,jackson2008social,girvan2002community,shmulevich2002probabilistic,bc09,bullmore09,rubinov10,carrington2005models,borgatti2009network,wasserman1994social,lazer2011networks}. However, a majority of methodological and applied studies have only considered interactions of one type among a set of entities of the same type. More recent studies have pointed to the heterogeneous and multimodal nature of such interactions, whereby a complex networked system is composed of multiple types of interactions among entities that themselves are of multiple types \citep{kivela14,boccaletti14,mucha10,pc15,paul2020spectral,sengupta2015spectral,sun2009ranking,lmw14,ferrara2014online,he2014comment,nickel2016review,paul2020random}.

Social relationships are known to affect individual behaviors and outcomes including dementia \citep{fratiglioni2000influence}, decision making \citep{kim2007impact}, adolescent smoking \cite{mercken2010dynamics}, and online behavior choices \citep{kwon2014social}. At the same time, individual attributes, such as race, age, and gender, can affect whether and how people form friendships or romantic partnerships \citep{dean2017friendship,mcpherson2001birds}. The effect of social relationships on individual behaviors is observed through disparities in the outcomes across different individuals when their friendship structures differ. The reciprocal is also observed through disparities in the friendship structures when individuals' attributes differ. Therefore, flexible joint modeling of social relationships and individual behaviors and attributes is needed to investigate their interrelationships effectively.

A number of popular models for social networks have been extended to incorporate nodal covariates in the literature, including, exponential random graph models (ERGMs) \citep{lusher2013exponential}, stochastic blockmodels \citep{mele2019spectral,sweet2015incorporating} and latent space models \citep{fosdick2015testing,krivitsky2008fitting,hoff2002latent,austin2013covariate}. In these models, the social network links are treated as dependent variables, and the effects of nodal covariates on the probability of network ties are subsequently estimated. Alternatively, social influence models use the node-level attributes as the dependent variables and estimate the effects of the social network on the attributes \citep{dorans1978alternative,leenders2002modeling,robins2001network,sweet2020latent,frank2004social,fujimoto2013decomposed,shalizi2011homophily,vanderweele2011sensitivity,vanderweele2013social,bramoulle2009identification,goldsmith2013social,shalizi2011homophily}. 

A different approach is to develop a joint modeling framework where different types of data are integrated by jointly modeling them as the dependent variables. In network science, the joint modeling framework has been proposed to model multi-view or multiplex networks, where multiple types of relations are observed on the same set of nodes, using stochastic block models \citep{barbillon2015stochastic,kefi2016structured}, and latent space models \citep{gollini2016joint,arroyo2019inference,salter2017latent,d2018node,zhang2020flexible}. In these models, latent variables are used to explain the probability of a node being connected with other nodes in multiple types of relations. When dependencies can be assumed across different layers, the common node representation is a flexible framework to summarize multiple types of information.

In many cases, high-dimensional multivariate covariates with complex latent structures are available in addition to a connected network.  In particular, often a social network is observed along with individuals' attributes or behavioral outcomes. In this case, two types of relations, the social network relations and various types of individual attributes, are observed among two types of nodes, the person nodes, and the attribute nodes. To jointly model such data, a multivariate normal distribution was fitted by \cite{fosdick2015testing} to the latent variables from the social network and the observed covariates. This work is in spirit the closest to our proposed model. However, it restricts the covariates to be normally distributed, and it does not take into account the multiple latent dimensions of the covariates. A dynamic version of \cite{fosdick2015testing} can be found in \cite{guhaniyogi2020joint} with possibilities to accommodate both categorical and continuous attributes. The most important distinction of our model from this line of work is that we use a second set of latent variables, the latent attribute variables, in addition to the latent person variables to summarize the information associated with each attribute. This modeling framework allows for the joint latent space, where two types of nodes are interactive instead of one. Other related works that jointly model heterogeneous networks are recently seen with stochastic block models including \cite{huang2020mixed,sengupta2015spectral}.

In this paper, we propose a joint latent space model for heterogeneous and multimodal networks with multiple types of relations among multiple types of nodes. The proposed Attribute Person Latent Space Model (APLSM) merges information from the social network and the multivariate covariates by assuming that the probabilities of a node being connected with other same-type and different-type nodes are explained by latent variables.  This model has a wide range of applications. For example in computer science, it is of interest to summarize relational data, e.g. likes and followers in social media with other user information such as personalities, health outcomes, online behavior choices, etc. In economics and business, it is of interest to summarize consumer information with their geographic networks and social networks. We demonstrate APLSM with a data set on the French financial elites \citep{kadushin1995friendship} available to download from \url{http://moreno.ss.uci.edu/data.html#ffe}. To fit the APLSM, we propose a Variational Bayesian Expectation-Maximization (VBEM) algorithm \cite{blei2017variational}. As an intermediate step we also develop a VBEM algorithm for fitting latent space models to bipartite networks. The variational methods enable the models to be fitted to large networks with high dimensional attributes, while our simulations show the accuracy of the methods.

The remainder of this paper is organized as follows. In section 2,  we introduce latent space models for bipartite networks and develop the variational inference approach to estimate the model. In section 3, we introduce the joint latent space model for the social network and the multivariate covariates and extend the variational inference method to the joint model. In section 4, we assess the performance of the estimators with a simulation study, and in section 5, we apply the proposed methodology to the French financial elite data. Finally, in section 6, we summarize our findings.

\section{Latent Space Models}

Development of Latent Space Models (LSM) for social networks can be traced back to \cite{hoff2002latent}'s latent distance and latent projection models. Both models assume that nodes can be positioned in a D-dimensional latent space and that the probability of an edge between two nodes depends on their closeness. The closer the two nodes, the less likely they form a connection. Conditional on these latent positions, the probability of two nodes forming an edge is independent of all the other edges in the network.  In the latent distance model, the Euclidean distances are used to describe the relationships between the nodes; whereas, in the latent projection model, scaled vector products are used to describe the relationships between nodes.

Let $\boldsymbol{Y_I}$ denote the $N \times N$ adjacency matrix of the social network among $N$ individuals. The $(i,j)$ th element of the matrix $\boldsymbol{Y_I}$, denoted as $y^{I}_{ij}$ is $1$ if person $i$ and person $j$ are related, for $i, j =\{1,2,\ldots,N\}$ and $i\neq j $.  Let $\boldsymbol{U}$ be a $N \times  D$ matrix of latent person position, each row of which is a $D$ dimensional vector $\boldsymbol{u_i} =(u_{i1},u_{i2},\ldots, u_{iD})$ indicating the latent position of person $i$ in the Euclidean space. The latent distance model for a binary social network $\boldsymbol{Y_I}$ can be written as: 
\begin{align*}
    Y^{I}_{ij}| (\boldsymbol{U}, \alpha_0) \sim Bernoulli( g(\phi_{ij})),
    \quad \quad g(\phi_{ij}) = \frac{\exp(\alpha_0 - | \boldsymbol{u}_i -\boldsymbol{u}_j |^2)}{1+\exp(\alpha_0 - | \boldsymbol{u}_i -\boldsymbol{u}_j |^2)}.
\end{align*}
We assume $\boldsymbol{u_i}$ to be the latent person variable,  $\boldsymbol{u_i}\overset{iid}{\sim} N(0,\lambda^2_0\textbf{I}_D)$, where $\alpha_0, \lambda^2_0$ are unknown parameters that need to be estimated and $\textbf{I}_D$ is the $D$ dimensional identity matrix. The probability of an edge increases as the Euclidean distance between the two nodes decreases. Here, we use the squared Euclidean distances $|\boldsymbol{u_i}-\boldsymbol{u_j}|^2$ instead of the Euclidean distances following \citep{gollini2016joint}. It has been shown in \cite{gollini2016joint} that squared Euclidean distances are computationally more efficient and that the latent positions obtained using squared Euclidean distances are extremely similar to those obtained using Euclidean distances.

Variations of the latent space model are further developed in \cite{hoff2005bilinear,hoff2008modeling,hoff2009multiplicative, hoff2018additive} and \cite{ma2020universal}. The extension of the latent distance model include \cite{hrt07}'s latent position cluster model that allows for the clustering of nodes based on the Euclidean distances. Replacing one Gaussian distribution with a mixture of Gaussians, \cite{hrt07} was able to account for the possible latent community structure in the networks. Additional random sender and receiver effects were added to the latent position cluster model by \cite{krivitsky2009representing}. The latent space model have been further extended to accommodate multiple networks \cite{gollini2016joint,salter2017latent}, dynamic networks \cite{sewell2015latent}, and bipartite networks \cite{friel2016interlocking}. The majority of the works on latent space models described above have utilized Bayesian estimation techniques including Markov Chain Monte Carlo (MCMC) and Variational Inference (VI). Recently, \cite{ma2020universal} has proposed two algorithms based on nuclear norm penalization and projected gradient descent to fit the latent space model with statistical consistency guarantees.

\subsection{Latent Space Model for Bipartite Networks}

Data on the attributes of the individuals in the network can be seen as a bipartite network with directed edges between two types of nodes. The development of the latent space model for bipartite networks includes a bipartite version of the latent cluster random effects model in the latentnet package \citep{krivitsky2008fitting}. In addition, the latent space model for a dynamic bipartite network was introduced by \citep{friel2016interlocking} to study the interlocking directorates in Irish companies. In the rest of this section, we introduce a Variational Bayesian EM algorithm for fitting a latent space model to binary bipartite networks (BLSM).

Let $\boldsymbol{Y_{IA}}$ denote the $N \times M$ bipartite network, whose $(i,a)$ th element $y^{IA}_{ia}$ is $1$ if person $i$ has attribute $a$, for $i =\{1,2,\ldots,N\}$ and $a =\{1,2,\ldots,M\}$. Let $\boldsymbol{V}$ be a $M \times  D$ matrix of latent attribute positions, each row of which is a $D$ dimensional vector $\boldsymbol{v_a} =(v_{a1},v_{a2}, \ldots, v_{aD})$ indicating the latent position of attribute $a$ in the Euclidean space.

The latent distance model for the bipartite network $\boldsymbol{Y_{IA}}$ can be written as: 
\begin{align*}
 Y^{IA}_{ia}| (\boldsymbol{U},\boldsymbol{V}, \alpha_0) \sim Bernoulli( g(\phi_{ia})),
    \quad \quad  g(\phi_{ia}) = \frac{\exp(\alpha_1 - | \boldsymbol{u}_i -\boldsymbol{v}_a |^2)}{1+\exp(\alpha_1 - | \boldsymbol{u}_i -\boldsymbol{v}_a |^2)},
    \label{bipar}
\end{align*} 
 We assume $\boldsymbol{u_i}\overset{iid}{\sim} N(0,\lambda^2_0\textbf{I}_D)$, $\boldsymbol{v_a}\overset{iid}{\sim} N(0,\lambda^2_1\textbf{I}_D)$, and $\alpha_1, \lambda_0$ and $\lambda_1$ to be unknown parameters. The parameter $\alpha_1$ accounts for the density of the bipartite network. The probability of a positive response increases as the Euclidean distance between the attribute node and the person node decreases.

\subsection{Variational Bayesian EM for the Bipartite Network}

We are interested in the posterior inference of the latent variables $\boldsymbol{u_i}$ and $\boldsymbol{v_a}$ following the distance  model conditioning on the observed bipartite network. The (conditional) posterior distribution is the ratio of the joint distribution of the observed data and unobserved latent variables to the observed data likelihood
\begin{equation*}
P(\boldsymbol{U},\boldsymbol{V} |\boldsymbol{Y_{IA}}) = \frac{ P(\boldsymbol{Y_{IA}}| \boldsymbol{U},\boldsymbol{V}) P(\boldsymbol{U},\boldsymbol{V})}{P(\boldsymbol{Y_{IA}})}.
\end{equation*}
We can characterize the distribution of latent positions and thus obtain the point and interval estimates by computing this posterior distribution. The variational inference algorithm is commonly used to estimate latent variables whose posterior distribution is intractable \citep{beal2003variational,attias1999inferring,beal2006variational,blei2017variational}. In network analysis, the variational approach has been proposed for the stochastic blockmodel \citep{dpr08,cdp11}, the mixed-membership stochastic
blockmodel \citep{airoldi2008mixed}, the multi-layer stochastic blockmodel \citep{xu14,pc15}, the dynamic stochastic blockmodel \citep{matias15}, the latent position cluster model \citep{salter2013variational} and the multiple network latent space model \citep{gollini2016joint}. Here, we propose a Variational Bayesian Expectation Maximization (VBEM) algorithm to approximate the posterior of the person and the attribute latent positions using the bipartite network. We propose a class of suitable variational posterior distributions for the conditional distribution of $(\boldsymbol{U},\boldsymbol{V} |\boldsymbol{Y_{IA}})$ and obtain a distribution from the class that minimizes the Kulback Leibler (KL) divergence from the true but intractable posterior.

We assign the following variational posterior distributions: $q(\boldsymbol{u_i})=N(\tilde{\textbf{u}}_i,\tilde{\Lambda}_{0})$ and $q(\boldsymbol{v_a})=N(\tilde{\textbf{v}}_a,\tilde{\Lambda}_{1})$ and set the joint distribution as
\begin{equation*}
q(\boldsymbol{U},\boldsymbol{V} | \boldsymbol{Y_{IA}})= \prod_{i=1}^Nq(\boldsymbol{u_i}) \prod_{a=1}^Mq(\boldsymbol{v_a}),
\end{equation*} 
where $\tilde{\textbf{u}}_i,\tilde{\Lambda}_{0},\tilde{\textbf{v}}_a,\tilde{\Lambda}_{1}$ are the parameters of the variational distribution, known as variational parameters.

We can estimate the variational parameters by minimizing the Kullback-Leiber (KL) divergence between the variational posterior $q(\boldsymbol{U},\boldsymbol{V}|\boldsymbol{Y_{IA}})$ and the true posterior $f(\boldsymbol{U},\boldsymbol{V}|\boldsymbol{Y_{IA}})$. Minimizing the KL divergence is equivalent to maximizing the following Evidence Lower Bound (ELBO) function \cite{blei2017variational}, (see detailed derivations in the Supplementary Materials)
\begin{align}
&\text{ELBO} = -\mathbb{E}_{q(\boldsymbol{U},\boldsymbol{V},\alpha_1| \boldsymbol{Y_{IA}})}\left[\frac{\log q(\boldsymbol{U},\boldsymbol{V},\alpha_1| \boldsymbol{Y_{IA}})}{\log p(\boldsymbol{U},\boldsymbol{V},\boldsymbol{Y_{IA}} |\alpha_1)}\right] \nonumber\\
&=-\int q(\boldsymbol{U},\boldsymbol{V},\alpha_1 |\boldsymbol{Y_{IA}})\log  \frac{q(\boldsymbol{U},\boldsymbol{V},\alpha_1 |\boldsymbol{Y_{IA}})}{f(\boldsymbol{U},\boldsymbol{V},\alpha_1 |\boldsymbol{Y_{IA}})}d(\boldsymbol{U},\boldsymbol{V},\alpha_1)\nonumber\\
&=-\int \prod_{i=1}^Nq(\boldsymbol{u_i}) \prod_{a=1}^Mq(\boldsymbol{v_a}) 
\log \frac{  \prod_{i=1}^Nq(\boldsymbol{u_i})  \prod_{a=1}^Mq(\boldsymbol{v_a})}{f(\boldsymbol{Y_{IA}} |\boldsymbol{U},\boldsymbol{V},\alpha_1) 
\prod_{i=1}^N f(\boldsymbol{u_i})\prod_{a=1}^M f(\boldsymbol{v_a})}d(\boldsymbol{U},\boldsymbol{V},\alpha_1)\nonumber\\
&= - \sum_{i=1}^N  \int q(\boldsymbol{u_i}) \log \frac{q(\boldsymbol{u_i})}{f(\boldsymbol{u_i})} d \boldsymbol{u_i} 
- \sum_{a=1}^M \int  q(\boldsymbol{v_a}) \log \frac{q(\boldsymbol{v_a})}{f(\boldsymbol{v_a})} d \boldsymbol{v_a}\nonumber \\
& \quad \quad + \int q(\boldsymbol{U},\boldsymbol{V},\alpha_1 |\boldsymbol{Y_{IA}})\log f(\boldsymbol{Y_{IA}} |\boldsymbol{U},\boldsymbol{V},\alpha_1)d(\boldsymbol{U},\boldsymbol{V},\alpha_1)\nonumber\\
&= - \sum_{i=1}^N\text{KL}[q(\boldsymbol{u_i})| f(\boldsymbol{u_i})]
 - \sum_{a=1}^M\text{KL}[q(\boldsymbol{v_a})| f(\boldsymbol{v_a})]\nonumber +  \mathbb{E}_{q(\boldsymbol{U},\boldsymbol{V},\alpha_1 |\boldsymbol{Y_{IA}})}[\log f(\boldsymbol{Y_{IA}} |\boldsymbol{U},\boldsymbol{V},\alpha_1)]\nonumber\\
& = - \frac{1}{2} \Big( DN \log (\lambda^2_0)- N\log (\det(\tilde{\Lambda}_{0})) \Big) - \frac{N \tr(\tilde{\Lambda}_0)}{2 \lambda^2_0}   - \frac{\sum_{i=1}^N \boldsymbol{\tilde{u}_i}^T\boldsymbol{\tilde{u}_i}}{2 \lambda^2_0} \nonumber\\
& \quad   - \frac{1}{2} \Big( DM \log (\lambda^2_1)- M\log (\det(\tilde{\Lambda}_{1})) \Big)
 -\frac{M \tr(\tilde{\Lambda}_1)}{2 \lambda^2_1} -\frac{\sum_{a=1}^M \boldsymbol{\tilde{v}_a}^T\boldsymbol{\tilde{v}_a}}{2 \lambda^2_1} + \frac{1}{2}(MD + ND) \nonumber\\ 
& \quad  + \mathbb{E}_{q(\boldsymbol{U},\boldsymbol{V} |\boldsymbol{Y_{IA}})}[\log f(\boldsymbol{Y_{IA}} |\boldsymbol{U},\boldsymbol{V})].  \label{kl1}
\end{align}
After applying Jensen's inequality \citep{jensen1906fonctions}, a lower-bound on $ \mathbb{E}_{q(\boldsymbol{U},\boldsymbol{V} |\boldsymbol{Y_{IA}})}[\log f(\boldsymbol{Y_{IA}} |\boldsymbol{U},\boldsymbol{V})]$ is given by,
\begin{align}
&\mathbb{E}_{q(\boldsymbol{U},\boldsymbol{V} |\boldsymbol{Y_{IA}})}[\log f(\boldsymbol{Y_{IA}} |\boldsymbol{U},\boldsymbol{V},\alpha_1)] \nonumber\\
\quad \quad \geq &  \sum_{i=1}^N \sum_{a=1}^M y_{ia}  \Bigg[ \tilde{\alpha}_1- \tr(\tilde{\Lambda}_{0} )- \tr(\tilde{\Lambda}_{1} )-(\boldsymbol{\tilde{u}_i}-\boldsymbol{\tilde{v}_a})^T(\boldsymbol{\tilde{u}_i}-\boldsymbol{\tilde{v}_a})   \Bigg] \nonumber\\
-&\sum_{i=1}^N \sum_{a=1}^M   \log \Bigg( 1+\frac{\exp(\tilde{\alpha}_1)}{ \det(\textbf{I} +2\tilde{\Lambda}_0+2\tilde{\Lambda}_1)^{\frac{1}{2}}}\exp \Big( -(\boldsymbol{\tilde{u}_i}-\boldsymbol{\tilde{v}_a})^T (\textbf{I} +2\tilde{\Lambda}_0+2\tilde{\Lambda}_1)^{-1}(\boldsymbol{\tilde{u}_i}-\boldsymbol{\tilde{v}_a})  \Big)  
\Bigg).\nonumber  \label{kl2}
\end{align}
We use the Variational Expectation-Maximization algorithm \citep{jordan1999introduction,baum1970maximization,dempster1977maximum} to maximize the ELBO function. Following the variational EM algorithm, we replace the E step of the celebrated EM algorithm, where we compute the expectation of the complete likelihood with respect to the conditional distribution $f(\boldsymbol{U},\boldsymbol{V}|\boldsymbol{Y_{IA}})$, with a VE step, where we compute the expectation with respect to the best variational distribution (obtained by optimizing the ELBO function) at that iteration. 

The detailed procedures are as follows. We start with the initial parameter, $\Theta^{(0)} = \tilde{\alpha}_1^{(0)},$ and $\tilde{\textbf{u}}_i^{(0)},\tilde{\Lambda}_{0}^{(0)},\tilde{\textbf{v}}_a^{(0)},\tilde{\Lambda}_{1}^{(0)}$, and then we iterate the following VE (Variational expectation) and M (maximization) steps. During the VE step, we maximize the $\text{ELBO}(q(\bf{U}), q(\bf{V}),\Theta)$ with respect to the variational parameters $\boldsymbol{\tilde{u}_i}, \boldsymbol{\tilde{v}_a}, \tilde{\lambda}_0$ and $\tilde{\lambda}_1$ given the other model parameters and obtain $\text{ELBO}(q^{*}(\bf{U}),q^{*}(\bf{V}),\Theta)$. During the M step, we fix $\boldsymbol{\tilde{u}_i}, \boldsymbol{\tilde{v}_a}, \tilde{\Lambda}_0$ and $\tilde{\Lambda}_1$ and maximize the $\text{ELBO}(q(\bf{U}), q(\bf{V}),\Theta)$ with respect to $\tilde{\alpha}_1$ . To do this, we differentiate $\text{ELBO}(q(\bf{U}), q(\bf{V}),\Theta)$ with respect to each variational parameter. We obtain closed form update rules by setting the partial derivatives to zero while introducing the first- and second-order Taylor series expansion approximation of the log functions in $\text{ELBO}(q(\bf{U}), q(\bf{V})$ $,\Theta)$ (see detailed derivations in supplementary material). The Taylor series expansions are commonly used in the variational approaches. For example, three first-order Taylor expansions were used by \cite{salter2013variational} to simplify the Euclidean distance in the latent position cluster model, and first- and second-order Taylor expansions were used by \cite{gollini2016joint} to simplify the squared Euclidean distance in LSM. Following the previous publications using Taylor expansions, we approximate the three log functions in our $\text{ELBO}(q(\bf{U}), q(\bf{V}),\Theta)$ function to find closed form update rules for the variational parameters. Define the function
\begin{equation*}
\boldsymbol{F_{ia}}=\sum_{i=1}^N \sum_{a=1}^M   \log \Bigg( 1+\frac{\exp(\tilde{\alpha}_1)}{ \det(\textbf{I} +2\tilde{\Lambda}_0+2\tilde{\Lambda}_1)^{\frac{1}{2}}}\exp \Big( -(\boldsymbol{\tilde{u}_i}-\boldsymbol{\tilde{v}_a})^T (\textbf{I} +2\tilde{\Lambda}_0+2\tilde{\Lambda}_1)^{-1}(\boldsymbol{\tilde{u}_i}-\boldsymbol{\tilde{v}_a})  \Big)  
\Bigg).
\end{equation*}
The closed form update rules of the ($t+1$)th iteration are as follows\\

\textbf{VE-step}: Estimate $\boldsymbol{\tilde{u}_i}^{(t + 1)}$, $\boldsymbol{\tilde{v}_a}^{(t + 1)}$, $\tilde{\Lambda}_0^{(t + 1)}$ and $\tilde{\Lambda}_1^{(t + 1)}$ by minimizing $\text{ELBO}(q(\bf{U}), q(\bf{V}),\Theta)$
\begin{align}
    & \boldsymbol{\tilde{u}_i^{(t+1)}}  =
    \Bigg[ \Bigg( \frac{1}{2 \lambda_0^2} + \sum_{a =1}^M y_{ia} ) \boldsymbol{I}  +  \frac{1}{2}  \boldsymbol{H_{ia}} (\boldsymbol{\tilde{u}_i^{(t)}} ) \Bigg]^{-1} \Bigg[ \sum_{a =1}^M y_{ia} \boldsymbol{\tilde{v}_a^{(t)}}   +    \frac{1}{2} \boldsymbol{H_{ia}} (\boldsymbol{\tilde{u}_i^{(t)}} )  \boldsymbol{\tilde{u}_i^{(t)}}   -  \frac{1}{2} \boldsymbol{G_{ia}} (\boldsymbol{\tilde{u}_i^{(t)}} ) \Bigg] \nonumber \\
& \boldsymbol{\tilde{v}_a^{(t+1)}} =
    \Bigg[ \Bigg( \frac{1}{2 \lambda_1^2} + \sum_{i =1}^N y_{ia}  \Bigg) \boldsymbol{I}  -  \frac{1}{2} \boldsymbol{H_{ia}} ( \boldsymbol{\tilde{v}_a^{(t)}}) \Bigg]^{-1} 
    \Bigg[  \sum_{i=1}^N y_{ia} \boldsymbol{\tilde{u}_i^{(t)}}    -  \frac{1}{2}\boldsymbol{G_{ia}} ( \boldsymbol{\tilde{v}_a^{(t)}}) \Bigg]\nonumber  \\
     &\tilde{\Lambda}_0^{(t+1)} = \frac{N}{2}
    \Bigg[ \Bigg( \frac{N}{2} \frac{1}{ \lambda_0^2}  + 
    \sum_{i =1}^N \sum_{a=1}^M y_{ia} \Bigg) \boldsymbol{I}  +  \boldsymbol{G_{ia}} (\tilde{\Lambda}_0^{(t)}) \Bigg]^{-1}&& \nonumber \\
    &\tilde{\Lambda}_1^{(t+1)} = \frac{M}{2}
    \Bigg[ \Bigg( \frac{M}{2}  \frac{1}{ \lambda_1^2} +  \sum_{i =1}^N \sum_{a=1}^M y_{ia}\Bigg) \boldsymbol{I}  + \boldsymbol{G_{ia}} (\tilde{\Lambda}_1^{(t)}) \Bigg]^{-1}, \label{b}
\end{align}where $\boldsymbol{G_{IA}}(\boldsymbol{\tilde{u}_i}^{(t)})$and $\boldsymbol{G_{IA}}(\boldsymbol{\tilde{v}_a}^{(t)})$ are the partial derivatives (gradients) of  $\boldsymbol{F_{IA}}$ with respect to $ \boldsymbol{\tilde{u}_i}$ and $\boldsymbol{\tilde{v}_a}$, evaluated at $ \boldsymbol{\tilde{u}_i}^{(t)}$ and $\boldsymbol{\tilde{v}_a}^{(t)}$, respectively. In $\boldsymbol{G_{IA}}(\boldsymbol{\tilde{u}_i}^{(t)})$, the subscript $\boldsymbol{IA}$ indicates that the gradient is of function $\boldsymbol{F_{IA}}$, and the subscript $i$ in $\boldsymbol{\tilde{u}_i}^{(t)}$ indicates that the gradient is with respect to $\boldsymbol{\tilde{u}_i}$, evaluated at $\boldsymbol{\tilde{u}_i}^{(t)}$. Similarly, $\boldsymbol{H_{IA}}(\boldsymbol{\tilde{u}_i}^{(t)})$ and $\boldsymbol{H_{IA}}(\boldsymbol{\tilde{v}_a}^{(t)})$ are the second-order partial derivatives of $ \boldsymbol{F_{IA}}$ with respect to $\boldsymbol{\tilde{u}_i}$ and $\boldsymbol{\tilde{v}_a}$, evaluated at $  \boldsymbol{\tilde{u}_i}^{(t)}$ and $\boldsymbol{\tilde{v}_a}^{(t)}$, respectively.\\ 

\textbf{M-step}: Estimate $\tilde{\alpha}_1^{(t + 1)}$ with the following closed form solution,
\begin{flalign}
    \tilde{\alpha}_1^{(t + 1)} =&\frac{\sum_{i=1}^N\sum_{a=1}^My_{ia}^{IA}- g_{IA}(\tilde{\alpha}_1^{(t)}) + \tilde{\alpha}_1^{(t)} h_{IA}(\tilde{\alpha}_1^{(t)} ) }{h_{IA}(\tilde{\alpha}_1^{(t)} )} \label{alpha1}, 
\end{flalign} where  $g_{IA}(\tilde{\alpha}_1^{(t)})$ is the partial derivative (gradient) of  $\boldsymbol{F_{IA}}$ with respect to  $\tilde{\alpha}_1$ , evaluated at $ \tilde{\alpha}_1^{(t)}$ ; and  $h_{IA}(\tilde{\alpha}_1^{(t)})$ is the second-order partial derivative of  $\boldsymbol{F_{IA}}$ with respect to $\tilde{\alpha}_1$ , evaluated at $ \tilde{\alpha}_1^{(t)}$.

\section{Joint Modeling of Social Network and Multivariate Attributes}

Next we define the attributes and persons latent space model (APLSM), which uses a single joint latent space to combine and summarize the information contained in both the person-person social network $\boldsymbol{Y_I}$, and person-attribute bipartite network formed by the high dimensional covariates $\boldsymbol{Y_{IA}}$. We assume that the persons and attributes can be positioned in an attribute-person joint latent space, which is a subset of the $D$ dimensional Euclidean space $\mathbb{R}^{D}$.  The latent person positions $\boldsymbol{U}$ is a shared latent variable that simultaneously affects both the social network and the multivariate covariates. In APLSM, we extend the conditional independence assumption of LSM by assuming that the probability of two nodes forming a connection in both matrices $\boldsymbol{Y_I}$ and $\boldsymbol{Y_{IA}}$ is independent of all other connections given the latent positions of the two nodes involved. In the APLSM, the joint distribution of the elements of the social network and the multivariate covariates can be written as 
\begin{align}
p(\boldsymbol{Y_I}, \boldsymbol{Y_{IA}}| \boldsymbol{U}, \boldsymbol{V}, \alpha_0, \alpha_1 ) & = \prod_{i =1}^N \prod_{j=1, j \neq i}^N p_1(y_{i,j}^I| \theta_{i,j}^I)  \prod_{i =1}^N \prod_{a=1}^M p_2(y_{i,a}^{IA}| \theta_{i,a}^{IA}), \nonumber \\
E(y_{i,j}^I | \theta_{i,j}^I)  = g_1(\theta_{i,j}^I),  \qquad
&  E(y_{i,a}^{IA} | \theta_{i,a}^{IA})  = g_2(\theta_{i,a}^{IA}), \nonumber 
\end{align}
where $g_i(\cdot)$ are the link functions, and $p_i(\cdot|\cdot)$ are the parametric families of distributions suitable for the type of data in the matrices. We set the priors $\boldsymbol{u_i}\overset{iid}{\sim} N(0,\lambda^2_0\textbf{I}_D)$, and $\boldsymbol{v_a}\overset{iid}{\sim} N(0,\lambda^2_1\textbf{I}_D)$. $\alpha_0, \alpha_1,  \lambda^2_0, \lambda^2_1$ are unknown parameters. Different levels of $\alpha_0$ and $\alpha_1$ account for different densities of the two networks. 
\begin{align}
    \theta_{i,j}^I  = \alpha_0-|\boldsymbol{u_i}-\boldsymbol{u_j}|^2, \qquad 
& \text{and} \qquad 
\theta_{i,a}^{IA}  = \alpha_1-|\boldsymbol{u_i}-\boldsymbol{v_a}|^2, \qquad \label{d}
\end{align}
 
In Equation \ref{d}, squared euclidean distance forms are assumed for $\theta_{i,j}^I$ and $\theta_{i,a}^{IA}$. If the data in $\boldsymbol{Y_I}$ and $\boldsymbol{Y_{IA}}$ are binary, then the link functions $g_1(\theta_{i,j}^I)$ and $g_2(\theta_{i,a}^{IA})$ are logistic inverse link functions, i.e., $g_1(\theta_{i,j}^I) = \frac{\exp(\theta_{i,j}^I)}{1+\exp(\theta_{i,j}^I)}$ and  $g_2(\theta_{i,a}^{IA})
 = \frac{\exp(\theta_{i,a}^{IA})}{1+\exp(\theta_{i,a}^{IA})}$, and the $p_i(\cdot|\cdot)$ are Bernoulli PDFs.

 In the APLSM, we define a attribute-person joint latent space, a hypothetical multidimensional space, in which the locations of the persons and the attributes follow predefined geometric rules reflecting each node's connection with another. While the latent person positions are more commonly seen in network science, the latent attribute positions describe the latent traits of the attributes seen through the persons' responses. The probability of person $i$ and person $j$ forming a friendly connection depends on the distance of $\boldsymbol{u_i}$ and $\boldsymbol{u_j}$ in the joint latent space. The smaller the latent distance between person $i$ and $j$, the higher the chance that person $i$ and person $j$ are friends. Similarly, the closer the latent positions of attribute $a$ and person $i$ are, the more likely that person $i$ shows attribute $a$. The relationships among persons also retain the transitivity and reciprocity properties: if person $i$ and $j$ form a bond, and person $i$ and $k$ are also friends, then person $j$ befriending person $i$ (reciprocity), and befriending person $k$ (transitivity) are both more likely. The transitivity property is preserved for relationships between persons and attributes: if persons $i$ and $j$ form a bond, and person $i$ indicate a presence for attribute $a$, then person $j$ indicating a presence for attribute $a$ is also more likely. 
 
While it is common for the edges in the friendship networks to be binary, the multivariate covariates can be more general. If the data in $\boldsymbol{Y_{IA}}$ are of discrete numerical scales, they can be modeled with other parametric families. For example, we can use $g_2(\theta_{i,a}^{IA})
 = \exp(\theta_{i,a}^{IA})$ as the Poisson inverse link function to model count data in $\boldsymbol{Y_{IA}}$, and thus $p_2(y_{i,a}^{IA}| \theta_{i,a}^{IA})$ becomes the PDF of the Poisson distribution. If the data in $\boldsymbol{Y_{IA}}$ are normally distributed, then the link function is the identity link. Alternatively, we can model the presence (or absence) of an edge separately from the weight of the edge (if it is present). For example, a zero inflated normal distribution was used by \cite{sewell2016latent} to model weighted edges, and the same goal was achieved by \cite{agarwal2019model} using a combination of Bernoulli distribution and a non-parametric weight distribution.

 In a similar fashion, the APLSM can be used to handle weighted edges. A zero inflated Poisson model for the distribution of $y_{i,a}^{IA}|\theta_{i,a}^{IA}$ can be seen as follows: 
\begin{align*}
p_3(y_{i,a}^{IA}|\theta_{i,a}^{IA}) & = (1-(\kappa(\theta_{i,a}^{IA}))^{(y_{i,a}^{IA}=0)} \times  \bigg \{(\kappa(\theta_{i,a}^{IA}) \prod \frac{\exp(-\gamma(\theta_{i,a}^{IA}))\gamma(\theta_{i,a}^{IA})^{y_{i,a}^{IA}}}{y_{i,a}^{IA!}}\bigg \} \\
\kappa(\theta_{i,a}^{IA}) & = \frac{  \exp(\theta_{i,a}^{IA})}{1+\exp(\theta_{i,a}^{IA})} \\
\gamma (\theta_{i,a}^{IA}) &= \exp(\theta_{i,a}^{IA}).
\end{align*}

\subsection*{The Variational Bayesian EM algorithm}

The VBEM algorithm for fitting APLSM can be derived following steps similar to section 2.2. We start by noting that the posterior distribution of APLSM is as follows:
\begin{equation*}
P(\boldsymbol{U},\boldsymbol{V} |\boldsymbol{Y_I},\boldsymbol{Y_{IA}}) = \frac{ P(\boldsymbol{Y_I},\boldsymbol{Y_{IA}}| \boldsymbol{U},\boldsymbol{V}) P(\boldsymbol{U},\boldsymbol{V})}{P(\boldsymbol{Y_I},\boldsymbol{Y_{IA}})}.
\end{equation*}
We assign the following variational posterior distribution: $q(\boldsymbol{u_i})=N(\tilde{\textbf{u}}_i,\tilde{\Lambda}_{0})$ and $q(\boldsymbol{v_a})=N(\tilde{\textbf{v}}_a,\tilde{\Lambda}_{1})$. We set the joint distribution as
\begin{equation*}
q(\boldsymbol{U},\boldsymbol{V} |\boldsymbol{Y_I},\boldsymbol{Y_{IA}})= \prod_{i=1}^Nq(\boldsymbol{u_i}) \prod_{a=1}^Mq(\boldsymbol{v_a}),
\end{equation*} where $\tilde{\textbf{u}}_i,\tilde{\Lambda}_{0},\tilde{\textbf{v}}_a,\tilde{\Lambda}_{1}$ are the parameters of the distribution, known as variational parameters.

The Evidence Lower Bound (ELBO) function for APLSM is (see detailed derivations in the Supplementary Materials)
\begin{align}
&\text{ELBO} = -\mathbb{E}_{q(\boldsymbol{U},\boldsymbol{V},\alpha_0,\alpha_1| \boldsymbol{Y_I},\boldsymbol{Y_{IA}})}[\frac{\log q(\boldsymbol{U},\boldsymbol{V},\alpha_0,\alpha_1| \boldsymbol{Y_I},\boldsymbol{Y_{IA}})}{\log p(\boldsymbol{U},\boldsymbol{V},\boldsymbol{Y_I},\boldsymbol{Y_{IA}} |\alpha_0,\alpha_1)}] \nonumber\\
&=-\int q(\boldsymbol{U},\boldsymbol{V},\alpha_0,\alpha_1 |\boldsymbol{Y_I},\boldsymbol{Y_{IA}})\log  \frac{q(\boldsymbol{U},\boldsymbol{V},\alpha_0,\alpha_1 |\boldsymbol{Y_I},\boldsymbol{Y_{IA}})}{f(\boldsymbol{U},\boldsymbol{V},\alpha_0,\alpha_1 |\boldsymbol{Y_I},\boldsymbol{Y_{IA}})}d(\boldsymbol{U},\boldsymbol{V},\alpha_0,\alpha_1)\nonumber\\
&=-\int \prod_{i=1}^Nq(\boldsymbol{u_i}) \prod_{a=1}^Mq(\boldsymbol{v_a}) 
\log \frac{  \prod_{i=1}^Nq(\boldsymbol{u_i})  \prod_{a=1}^Mq(\boldsymbol{v_a})}{f(\boldsymbol{Y_I},\boldsymbol{Y_{IA}} |\boldsymbol{U},\boldsymbol{V},\alpha_0,\alpha_1) 
\prod_{i=1}^N f(\boldsymbol{u_i})\prod_{a=1}^M f(\boldsymbol{v_a})}d(\boldsymbol{U},\boldsymbol{V},\alpha_0,\alpha_1)\nonumber\\
&= - \sum_{i=1}^N  \int q(\boldsymbol{u_i}) \log \frac{q(\boldsymbol{u_i})}{f(\boldsymbol{u_i})} d \boldsymbol{u_i} 
- \sum_{a=1}^M \int  q(\boldsymbol{v_a}) \log \frac{q(\boldsymbol{v_a})}{f(\boldsymbol{v_a})} d \boldsymbol{v_a}\nonumber \\
& \quad \quad + \int q(\boldsymbol{U},\boldsymbol{V},\alpha_0,\alpha_1 |\boldsymbol{Y_I},\boldsymbol{Y_{IA}})\log f(\boldsymbol{Y_I},\boldsymbol{Y_{IA}} |\boldsymbol{U},\boldsymbol{V},\alpha_0,\alpha_1)d(\boldsymbol{U},\boldsymbol{V},\alpha_0,\alpha_1)\nonumber\\
&= - \sum_{i=1}^N\text{KL}[q(\boldsymbol{u_i})| f(\boldsymbol{u_i})]
 - \sum_{a=1}^M\text{KL}[q(\boldsymbol{v_a})| f(\boldsymbol{v_a})] + \mathbb{E}_{q(\boldsymbol{U},\boldsymbol{V},\alpha_0,\alpha_1 |\boldsymbol{Y_I},\boldsymbol{Y_{IA}})}[\log f(\boldsymbol{Y_I},\boldsymbol{Y_{IA}} |\boldsymbol{U},\boldsymbol{V},\alpha_0,\alpha_1)]\nonumber\\
& = - \frac{1}{2} \Big( DN \log (\lambda^2_0)- N\log (\det(\tilde{\Lambda}_{0})) \Big) - \frac{N \tr(\tilde{\Lambda}_0)}{2 \lambda^2_0}   - \frac{\sum_{i=1}^N \boldsymbol{\tilde{u}_i}^T\boldsymbol{\tilde{u}_i}}{2 \lambda^2_0} \nonumber\\
& \quad   - \frac{1}{2} \Big( DM \log (\lambda^2_1)- M\log (\det(\tilde{\Lambda}_{1})) \Big)
 -\frac{M \tr(\tilde{\Lambda}_1)}{2 \lambda^2_1} -\frac{\sum_{a=1}^M \boldsymbol{\tilde{v}_a}^T\boldsymbol{\tilde{v}_a}}{2 \lambda^2_1} + \frac{1}{2}(MD + ND) \nonumber\\ 
& \quad  + \mathbb{E}_{q(\boldsymbol{U},\boldsymbol{V} |\boldsymbol{Y_I},\boldsymbol{Y_{IA}})}[\log f(\boldsymbol{Y_I},\boldsymbol{Y_{IA}} |\boldsymbol{U},\boldsymbol{V})]  \label{kl1}. 
\end{align}
After applying Jensen's inequality \citep{jensen1906fonctions}, a lower-bound on the third term is given by,
\begin{align}
&\mathbb{E}_{q(\boldsymbol{U},\boldsymbol{V} |\boldsymbol{Y_I},\boldsymbol{Y_{IA}})}[\log f(\boldsymbol{Y_I},\boldsymbol{Y_{IA}} |\boldsymbol{U},\boldsymbol{V},\alpha_0,\alpha_1)] \nonumber\\
\quad \quad \geq &  \sum_{i=1}^N \sum_{a=1}^M y_{ia}  \Bigg[ \tilde{\alpha}_1- \tr(\tilde{\Lambda}_{0} )- \tr(\tilde{\Lambda}_{1} )-(\boldsymbol{\tilde{u}_i}-\boldsymbol{\tilde{v}_a})^T(\boldsymbol{\tilde{u}_i}-\boldsymbol{\tilde{v}_a})   \Bigg] \nonumber\\
+&\sum_{i=1}^N \sum_{j=1, j \neq i}^N y_{ij}  \Bigg[ \tilde{\alpha}_0-2 \tr(\tilde{\Lambda}_{0} )- (\boldsymbol{\tilde{u}_i}-\boldsymbol{\tilde{u}_j})^T(\boldsymbol{\tilde{u}_i}-\boldsymbol{\tilde{u}_j})   \Bigg]\nonumber\\
-&\sum_{i=1}^N \sum_{a=1}^M   \log \Bigg( 1+\frac{\exp(\tilde{\alpha}_1)}{ \det(\textbf{I} +2\tilde{\Lambda}_0+2\tilde{\Lambda}_1)^{\frac{1}{2}}}\exp \Big( -(\boldsymbol{\tilde{u}_i}-\boldsymbol{\tilde{v}_a})^T (\textbf{I} +2\tilde{\Lambda}_0+2\tilde{\Lambda}_1)^{-1}(\boldsymbol{\tilde{u}_i}-\boldsymbol{\tilde{v}_a})  \Big)  
\Bigg)\nonumber\\
-&\sum_{i=1}^N \sum_{j=1, j \neq i}^N \log \Bigg( 1+ \frac{\exp(\tilde{\alpha}_0)}{\det (\textbf{I} + 4\tilde{\Lambda}_{0})^{1/2}} \exp \Big( -(\boldsymbol{\tilde{u}_i}-\boldsymbol{\tilde{u}_j})^T (\textbf{I} + 4\tilde{\Lambda}_{0})^{-1}(\boldsymbol{\tilde{u}_i}-\boldsymbol{\tilde{u}_j})  \Big)  \Bigg)  \label{kl2}. 
\end{align}
In addition to $\boldsymbol{F_{ia}}$, we also take into account  $\boldsymbol{F_{i}}$
\begin{align*}
\begin{split}
&\boldsymbol{F_i}=\sum_{i=1}^N \sum_{j=1, j \neq i}^N \log \Bigg( 1+ \frac{\exp(\tilde{\alpha}_0)}{\det (\textbf{I} + 4\tilde{\Lambda}_{0})^{1/2}} \exp \Big( -(\boldsymbol{\tilde{u}_i}-\boldsymbol{\tilde{u}_j})^T (\textbf{I} + 4\tilde{\Lambda}_{0})^{-1}(\boldsymbol{\tilde{u}_i}-\boldsymbol{\tilde{u}_j})  \Big)  \Bigg). 
\end{split}
\end{align*}
The closed form update rules of the ($t+1$)th iteration are as follows\\

\textbf{VE-step}: Estimate $\boldsymbol{\tilde{u}_i}^{(t + 1)}$, $\boldsymbol{\tilde{v}_a}^{(t + 1)}$, $\tilde{\Lambda}_0^{(t + 1)}$ and $\tilde{\Lambda}_1^{(t + 1)}$ by minimizing $\text{ELBO}(q(\bf{U}), q(\bf{V}),\Theta)$
\begin{align}
    & \boldsymbol{\tilde{u}_i^{(t+1)}}  =
    \Bigg[ \Bigg( \frac{1}{2 \lambda_0^2} + \sum_{j =1, j \neq i}^N  ( y_{ij} +  y_{ji} )+ \sum_{a =1}^M y_{ia} ) \boldsymbol{I} + \boldsymbol{H_i} (\boldsymbol{\tilde{u}_i^{(t)}} ) +  \frac{1}{2}  \boldsymbol{H_{ia}} (\boldsymbol{\tilde{u}_i^{(t)}} ) \Bigg]^{-1} \label{u} \\
    &\Bigg[ \sum_{j =1, j \neq i}^N ( y_{ij} +  y_{ji} )\boldsymbol{\tilde{u}_j} +  \sum_{a =1}^M y_{ia} \boldsymbol{\tilde{v}_a^{(t)}}   - \boldsymbol{G_i} (\boldsymbol{\tilde{u}_i^{(t)}} ) +  \Big( \boldsymbol{H_i} (\boldsymbol{\tilde{u}_i^{(t)}} ) +  \frac{1}{2} \boldsymbol{H_{ia}} (\boldsymbol{\tilde{u}_i^{(t)}} ) \Big) \boldsymbol{\tilde{u}_i^{(t)}}   -  \frac{1}{2} \boldsymbol{G_{ia}} (\boldsymbol{\tilde{u}_i^{(t)}} ) \Bigg] \label{v}\\
& \boldsymbol{\tilde{v}_a^{(t+1)}} =
    \Bigg[ \Bigg( \frac{1}{2 \lambda_1^2} + \sum_{i =1}^N y_{ia}  \Bigg) \boldsymbol{I}  -  \frac{1}{2} \boldsymbol{H_{ia}} ( \boldsymbol{\tilde{v}_a^{(t)}}) \Bigg]^{-1} 
    \Bigg[  \sum_{i=1}^N y_{ia} \boldsymbol{\tilde{u}_i^{(t)}}    -  \frac{1}{2}\boldsymbol{G_{ia}} ( \boldsymbol{\tilde{v}_a^{(t)}}) \Bigg]\nonumber \\
     &\tilde{\Lambda}_0^{(t+1)} = \frac{N}{2}
    \Bigg[ \Bigg( \frac{N}{2} \frac{1}{ \lambda_0^2} + 2\sum_{i =1}^N \sum_{j =1}^N y_{ij} + 
    \sum_{i =1}^N \sum_{a=1}^M y_{ia} \Bigg) \boldsymbol{I} +  \boldsymbol{G_i} (\tilde{\Lambda}_0^{(t)}) +  \boldsymbol{G_{ia}} (\tilde{\Lambda}_0^{(t)}) \Bigg]^{-1}&& \label{lam0}\\
    &\tilde{\Lambda}_1^{(t+1)} = \frac{M}{2}
    \Bigg[ \Bigg( \frac{M}{2}  \frac{1}{ \lambda_1^2} +  \sum_{i =1}^N \sum_{a=1}^M y_{ia}\Bigg) \boldsymbol{I}  + \boldsymbol{G_{ia}} (\tilde{\Lambda}_1^{(t)}) \Bigg]^{-1}, 
    \label{lam1}
\end{align} 
where $\boldsymbol{G_I}(\boldsymbol{\tilde{u}_i}^{(t)})$, $\boldsymbol{G_{IA}}(\boldsymbol{\tilde{u}_i}^{(t)})$and $\boldsymbol{G_{IA}}(\boldsymbol{\tilde{v}_a}^{(t)})$ are the partial derivatives (gradients) of $\boldsymbol{F_I},  \boldsymbol{F_{IA}}$ and $\boldsymbol{F_{IA}}$ with respect to $\boldsymbol{\tilde{u}_i},  \boldsymbol{\tilde{u}_i}$ and $\boldsymbol{\tilde{v}_a}$, evaluated at $\boldsymbol{\tilde{u}_i}^{(t)}, \boldsymbol{\tilde{u}_i}^{(t)}$ and $\boldsymbol{\tilde{v}_a}^{(t)}$, respectively. In $\boldsymbol{G_I}(\boldsymbol{\tilde{u}_i}^{(t)})$, the subscript $I$ indicates that the gradient is of function $\boldsymbol{F_I}$, and the subscript $i$ in $\boldsymbol{\tilde{u}_i}^{(t)}$ indicates that the gradient is with respect to $\boldsymbol{\tilde{u}_i}$, evaluated at $\boldsymbol{\tilde{u}_i}^{(t)}$. Similarly, $\boldsymbol{H_I}(\boldsymbol{\tilde{u}_i}^{(t)})$,  $\boldsymbol{H_{IA}}(\boldsymbol{\tilde{u}_i}^{(t)})$ and $\boldsymbol{H_{IA}}(\boldsymbol{\tilde{v}_a}^{(t)})$ are the second-order partial derivatives of $\boldsymbol{F_I}, \boldsymbol{F_{IA}}$ and $\boldsymbol{F_{IA}}$ with respect to $\boldsymbol{\tilde{u}_i}, \boldsymbol{\tilde{v}_a}, \boldsymbol{\tilde{u}_i}$ and $\boldsymbol{\tilde{v}_a}$, evaluated at $\boldsymbol{\tilde{u}_i}^{(t)}, \boldsymbol{\tilde{v}_a}^{(t)}, \boldsymbol{\tilde{u}_i}^{(t)}$ and $\boldsymbol{\tilde{v}_a}^{(t)}$, respectively.\\ 

\textbf{M-step}: Estimate $\tilde{\alpha}_0^{(t + 1)}$ and $\tilde{\alpha}_1^{(t + 1)}$ with the following update rules,
\begin{flalign}
     \tilde{\alpha}_0^{(t + 1)} =&\frac{\sum^N_{i=1}\sum^N_{j=1}y_{ij}^I- g_I(\tilde{\alpha}_0^{(t)}) + \tilde{\alpha}_0^{(t)} h_I(\tilde{\alpha}_0^{(t)} ) }{h_I(\tilde{\alpha}_0^{(t)} )}&& \label{alpha0}\\
    \tilde{\alpha}_1^{(t + 1)} =&\frac{\sum_{i=1}^N\sum_{a=1}^My_{ia}^{IA}- g_{IA}(\tilde{\alpha}_1^{(t)}) + \tilde{\alpha}_1^{(t)} h_{IA}(\tilde{\alpha}_1^{(t)} ) }{h_{IA}(\tilde{\alpha}_1^{(t)} )},  \label{alpha1}
\end{flalign} where $g_I(\tilde{\alpha}_0^{(t)})$ and $g_{IA}(\tilde{\alpha}_1^{(t)})$ are the partial derivatives (gradients) of $\boldsymbol{F_I}$ and $\boldsymbol{F_{IA}}$ with respect to $\tilde{\alpha}_0$ and $\tilde{\alpha}_1$, evaluated at $\tilde{\alpha}_0^{(t)}, \tilde{\alpha}_1^{(t)}$; and $h_I(\tilde{\alpha}_0^{(t)})$ and $h_{IA}(\tilde{\alpha}_1^{(t)})$ are the second-order partial derivatives of $\boldsymbol{F_I}$ and $\boldsymbol{F_{IA}}$ with respect to $\tilde{\alpha}_0$ and $\tilde{\alpha}_1$, evaluated at $\tilde{\alpha}_0^{(t)}, \tilde{\alpha}_1^{(t)}$.

\begin{algorithm}[tb]
  \caption{VBEM Estimation procedure}
  \label{alg:estimation}
  
  \begin{algorithmic}
    \STATE {\bfseries Input:} Network Adjacency matrix $Y_I$, Multivariate Covariates matrix $Y_{IA}$, number of dimensions $D$
    \STATE {\bfseries Result:} Model parameters $\Theta = \tilde{\alpha}_0,\tilde{\alpha}_1,$ and $\tilde{\textbf{u}}_i,\tilde{\Lambda}_{0},\tilde{\textbf{v}}_a,\tilde{\Lambda}_{1}$
  \end{algorithmic}

  \begin{algorithmic}[1]
    
    \WHILE{$t < N_{\text{iter}}$ and $\text{KL}_{\text{dis}} < .999999$} \label{line:estimationForLoop}
         \STATE Compute estimates $\tilde{\alpha}_0$ and     $\tilde{\alpha}_1$ using Equations \ref{alpha0} and \ref{alpha1}
    \STATE Compute estimates $\tilde{\Lambda}_{0}$ and $\tilde{\Lambda}_{1}$ using Equations \ref{lam0} and \ref{lam1}
     \STATE Compute estimates $\tilde{\textbf{u}}_i$ and $\tilde{\textbf{v}}_a$ using Equations \ref{u} and \ref{v}
     
     \STATE Compute estimates KL using Equations \ref{kl1} and \ref{kl2}
     
      \STATE $\text{KL}_{\text{dis}} \leftarrow$ $\text{KL}^t/\text{KL}^{t-1}$
       \STATE $t \leftarrow$ $t +1$

    \ENDWHILE

    \STATE \textbf{return} $[ \tilde{\alpha}_0,\tilde{\alpha}_1,$, $\tilde{\textbf{u}}_i,\tilde{\Lambda}_{0},\tilde{\textbf{v}}_a,\tilde{\Lambda}_{1},\text{KL}]$
  \end{algorithmic}
\end{algorithm}

\section{Simulation Study}

\begin{figure}
\begin{center}
    \includegraphics[width=1\textwidth]{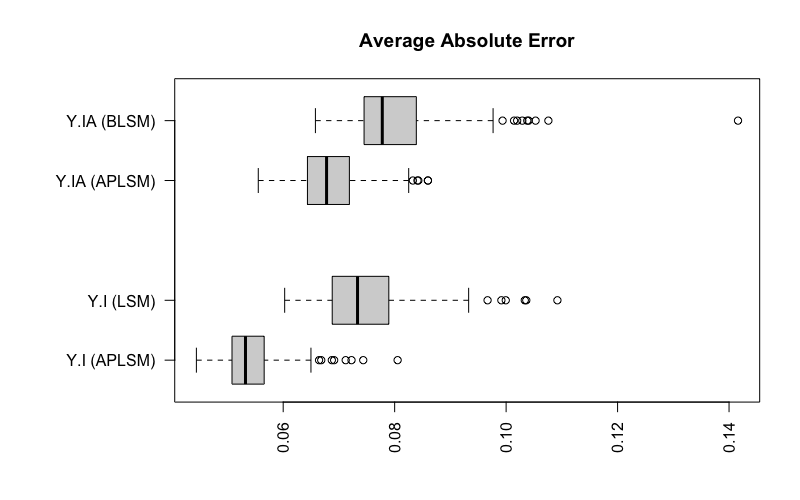}
  \end{center}
  \vspace{-25pt}
      \caption{The distributions of the AAEs when $\alpha_0 = 2$ and $\alpha_1 = 1.5$.}
\label{aae}
\end{figure}

\begin{figure}[h]
\begin{center}
\begin{subfigure}{0.5\textwidth}
    \includegraphics[width=1\textwidth]{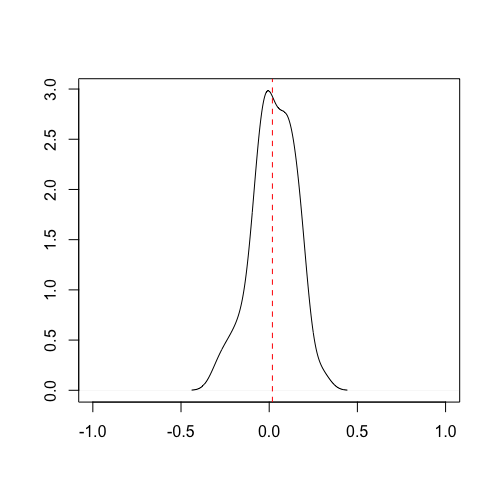}
  \end{subfigure}%
  \begin{subfigure}{0.5\textwidth}
\includegraphics[width=1\textwidth]{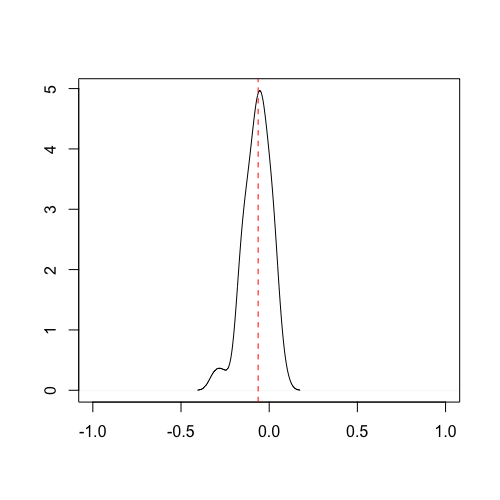}
\end{subfigure}
  \end{center}
   \caption{ Distributions of the distances between the true and estimated $\alpha_0$ (left), $\alpha_1$ (right) and when $\alpha_0 = 0.5$ and $\alpha_1 = 0$. }\label{alpha}
\end{figure}

\begin{figure}[h]
\begin{center}
\begin{subfigure}{0.5\textwidth}
    \includegraphics[width=1\textwidth]{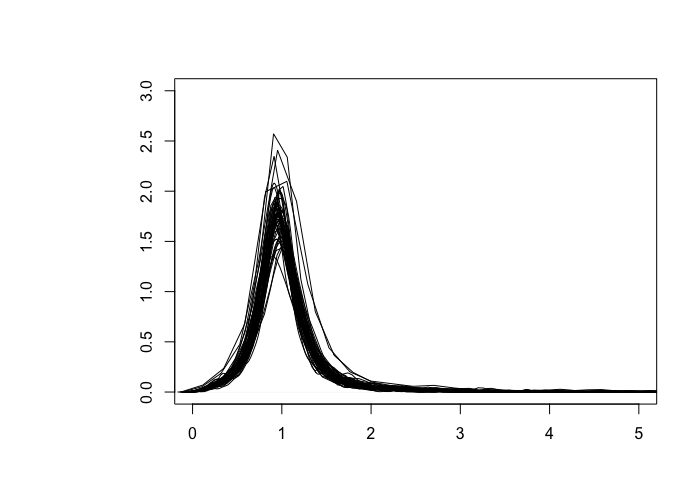}
  \end{subfigure}%
  \begin{subfigure}{0.5\textwidth}
\includegraphics[width=1\textwidth]{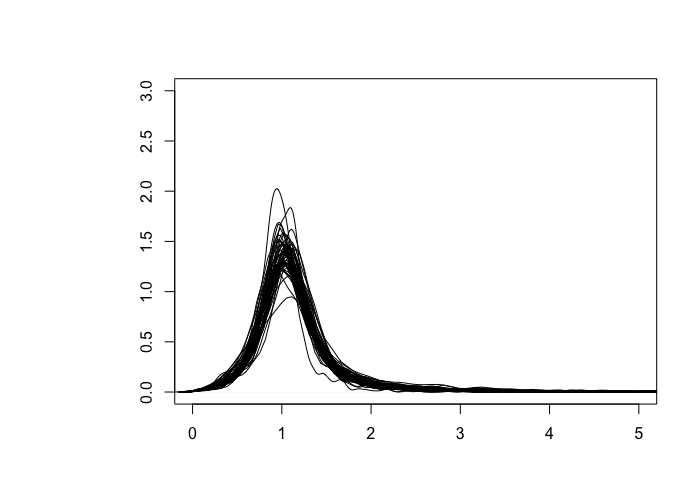}
\end{subfigure}
  \end{center}
      \caption{ Distributions of pairwise distance ratios, comparing $\boldsymbol{\tilde{u}_i}$ with $\boldsymbol{u_i}$ (left) and $\boldsymbol{\tilde{v}_a}$ with $\boldsymbol{v_a}$ (right) when $\alpha_0 = -1$ and $\alpha_1 = 0.5$.}\label{Z}
\end{figure}

In this section, we conduct a simulation study to evaluate the proposed VBEM algorithm's performance for the APLSM. We compare APLSM with two baseline approaches, namely LSM using variational inference proposed by \cite{gollini2016joint}, and LSM for bipartite networks (BLSM) using variational inference developed in Section 2.2. We expect that with the inclusion of the information in $\boldsymbol{Y_{IA}}$, the APLSM will exhibit a stronger model recovery fit than the LSM for the social network. Similarly, we expect the APLSM to be more successful in model recovery than the BLSM due to the addition of information from $\boldsymbol{Y_{I}}$.

To assess whether the proposed algorithm can recover the true link probabilities, we use the average absolute error (AAE) between the true link probabilities and the estimated link probabilities as a metric. The smaller the AAEs, the closer the estimated link probabilities are to the true link probabilities, indicating better model recovery. To assess whether the proposed algorithm can recover true $\alpha$ values, we look at the differences between the estimated values and the true values. If the differences are concentrated around mean 0 with a small variance, it will indicate a good recovery. Finally, we assess the fit of the estimated latent positions by calculating the proportions of the pairwise distances based on the estimated latent positions to the pairwise distances based on the true latent positions. Even though the true latent positions cannot be exactly recovered in any latent space model due to unidentifiability, we expect a successful algorithm to be able to recover the true distances \cite{sewell2015latent}. If the estimated latent positions preserve the nodes' relative positions in the latent space, we expect the proportion of the estimated pairwise distances to the true pairwise distances to be close to 1.

The design of the simulation is as follows. We first generate data following Equation \ref{d}. To do this, we set $\lambda_0$ and $\lambda_1$ to be $1$ and the number of attributes and persons to be 50. We sample $\boldsymbol{Z_I}$ and $\boldsymbol{Z_A}$ from the multivariate normal distributions using the above parameter values. We produce true link probabilities between persons and between attributes and persons using different sets of $\alpha$ values. Different $\alpha$ values are associated with different densities of the data. Then, we generate $\boldsymbol{Y_I}$ and $\boldsymbol{Y_{IA}}$ matrices. Each entry of the matrices is independently generated from the Bernoulli distribution using the corresponding link probability. We apply the APLSM with the VBEM estimator to the generated data and obtain the latent positions' posterior distributions and estimates for the fixed parameters. We use the posterior means as the point estimates of the latent positions to obtain the estimated probabilities. We also fit LSM to $\boldsymbol{Y_I}$ and BLSM to $\boldsymbol{Y_{IA}}$. Using the posterior means as point estimates of the latent positions, we obtain the estimated probabilities for $\boldsymbol{Y_I}$ and $\boldsymbol{Y_{IA}}$, respectively. We repeat this process $200$ times and report the results.

In Figure \ref{aae}, we present the boxplots over 200 simulations of averages of the absolute differences (AAEs) across entries in $\boldsymbol{Y_I}$ and $\boldsymbol{Y_{IA}}$ for APLSM, LSM and BLSM when $\alpha_0 = 2$ and $\alpha_1 = 1.5$. We see that the AAEs are generally smaller when an APLSM is fitted to the data compared to AAEs when an LSM is fitted. Therefore there is a strong improvement in model recovery using APLSM for $\boldsymbol{Y_I}$ compared with using LSM. Clearly, this improvement results from the added attribute information that consolidates the latent person position estimates. Similarly, the AAEs are smaller based on fitting the APLSM than AAEs based on fitting the BLSM. Again, this suggests a big improvement in model recovery using APLSM for $\boldsymbol{Y_{IA}}$ than using BLSM. In this case, the added social network consolidates the latent person position estimates resulting in a better model recovery for APLSM.

In Figure \ref{alpha}, we present the distributions of the differences between the estimated $\alpha$ values and the true $\alpha$ values. The true $\alpha$ values are $.5$ and $0$ for $\alpha_0$ and $\alpha_1$. The distribution of the differences for each $\alpha$ is centered around $0$, indicating little bias in the $\alpha$ estimates. Both also are relatively narrow, implying that the estimated $\alpha$ values are precise and close to the true $\alpha$ values.

Finally, in Figure \ref{Z}, we compare the pairwise distances based on the estimated latent positions with those based on the true latent positions. The distance between nodes $i$ and $j$ using estimated $\boldsymbol{\tilde{u}_i}$ and $\boldsymbol{\tilde{u}_j}$ should be close to the true distance between nodes $i$ and $j$ using $\boldsymbol{u_i}$ and $\boldsymbol{u_j}$ if the VBEM estimation algorithm successfully maintains and recovers the relationship between nodes $i$ and $j$. As shown in both plots in Figure \ref{Z}, the distributions of the ratios of the estimated and true pairwise distances are narrow and centered around 1, implying satisfactory recoveries of the nodes' relationships with each other through the estimated latent positions.

\section{Application}

We apply APLSM to the French financial elite dataset collected by Kadushin and de Quillacqi to study the friendship network among top financial elites in France during the last years of the Socialist government \cite{kadushin1995friendship}. We first introduce the data and then describe our results in detail.

\subsection*{Data}
The data were collected through interviews for people who held leading positions in major financial institutions and frequently appeared in financial section press reports. The friendship information was collected by asking the interviewees to name their friends in the social context. Kadushin and de Quillacqi then identified an inner circle of 28 elites from the initial sample based on their influence and their perceived eliteness by other participants. The resulting friendship network is a symmetric adjacency matrix. 

The data also contains additional background information, including age, a complex set of post-secondary education experiences, place of birth, political and cabinet careers, political party preference, religion, current residence, and club memberships. Two aspects of the elites' "prestige" include whether the person is named in the social register and whether the person has a \textit{particle} ("de") in front of either his (no woman was in the inner circle), his wife's, his mother's, or his children's names. Having "de" in the name is associated with nobility. Father's occupation is one of the variables used to reflect an elite's social class. Fathers' occupation is considered ``high'' if the father is in higher management, a professional, an industrialist, or an investor. Unfortunately, upon communications with the original author, we found that the coding procedures regarding some variables have been lost, including Finance Ministry information, religion, etc. We end up with $13$ binary variables including information on education (``Science Po'', ``Polytechniqu'', ``University'' and ``Ecole Nationale d'Administration''), career (``Inspection General de Finance'' and ``Cabinet''), class (``Social Register'', ``Father Status'', ``Particule''), politics (``Socialist'', ``Capitalist'' and ``Centrist'') and ``Age'' after excluding the lost or the unrelated information, i.e.,  mason and location, which are not associated with the social network based on \cite{kadushin1995friendship} (location is not considered to be related to the social network after adjusting for multiple comparisons). ``Age'' was converted into a binary variable following \cite{kadushin1995friendship}, where a group of elites was considered of older age with an average birth year of 1938. We will use ENA as an abbreviation for Ecole Nationale d'Administration. 

The science Po and the other educational variables warrant further explanations. The Science Po or the Institut d'Etudes Politiques de Paris prepares students for the entrance exam of the 
ENA. An alternative of the Science Po is the (Ecole) Polytechnique, a French military school whose graduates often enter one of the technical ministries. These elites with Polytechnique degrees enter one of the technical ministries. Both the Science Po and the Polytechnique are called Grandes Ecoles. A Grandes Ecoles education is highly respected in France as it leads to membership in the ENA, where the grands corps, which are the French civil service elites, including the Inspection General de Finance, etc-recruit its members \citep{kadushin1995friendship}.

\subsection*{Previous Approaches}

The authors in \cite{kadushin1995friendship} first used multidimensional scaling to draw the friendship network's sociogram. Then they applied Quadratic Assignment Procedure regressions and correlations to test each background variable's association with the social network. Based on the social network, two clusters were identified, which the authors called the left and the right moieties. The dependence between the social network and background information was understood through comparisons of the elites between the left and the right moiety. The elites in the right moiety were found to have a higher social class (upper-class parentage with high social standing), to be older (average birth year of 1929), and to have fewer appointments in public offices. The left moiety elites were more likely to be ENA graduates, grand corps members, cabinet members, treasury service members, socialists, and younger (average birth year of 1938). 

Using the APLSM, we will construct a joint latent space which will allow us to jointly model elites' friendship connections and their background information. Using the APLSM, we will also replicate \cite{kadushin1995friendship}'s left and the right moiety, adding simultaneous interpretation for the division in the elite circle. Furthermore, we observe an additional division within the left moiety using APLSM, which provides opportunities for new hypotheses.

\subsection*{Analyses and Findings}

We fit the APLSM model to the social network $\boldsymbol{Y_I}$ and the covariates $\boldsymbol{Y_{IA}}$ of the French financial elites dataset. For comparison, we also fit the social network $\boldsymbol{Y_I}$ and the covariates $\boldsymbol{Y_{IA}}$ separately using LSM. We fit the latent space model to the friendship network using the variational inference proposed in \cite{gollini2016joint}. For the multivariate covariates, we fit the bipartite latent space model (BLSM) using the variational inference method we developed in Section 2.2.

\begin{figure}
\begin{center}
    \includegraphics[width=1\textwidth]{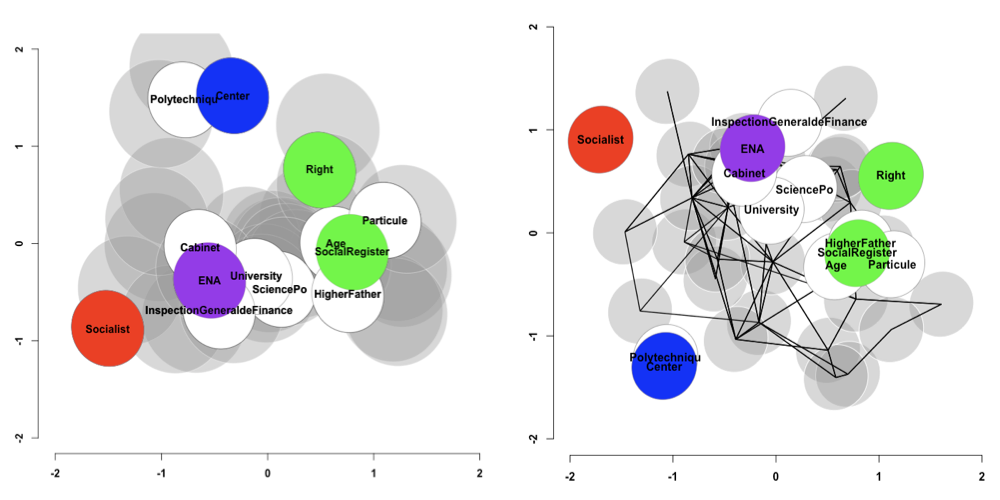}
  \end{center}
  \vspace{-25pt}
      \caption{The estimated latent person and attribute positions, $\boldsymbol{\tilde{u}_i}$ and $\boldsymbol{\tilde{v}_a}$ based on the BLSM (left) and the APLSM (right). The white ellipses represent $80\%$ approximate credible intervals for the $\tilde{\textbf{v}}_a$, and the grey ellipses represent $80\%$ approximate credible intervals for the $\tilde{\textbf{u}}_i$. The latent positions of the attributes Center, Right, Social register, ENA and Socialist are colored as blue, green, green, purple and red.  The black edges represent the friendship edges.}
\label{total}
\end{figure}

\begin{figure}
\begin{center}
    \includegraphics[width=0.95\textwidth]{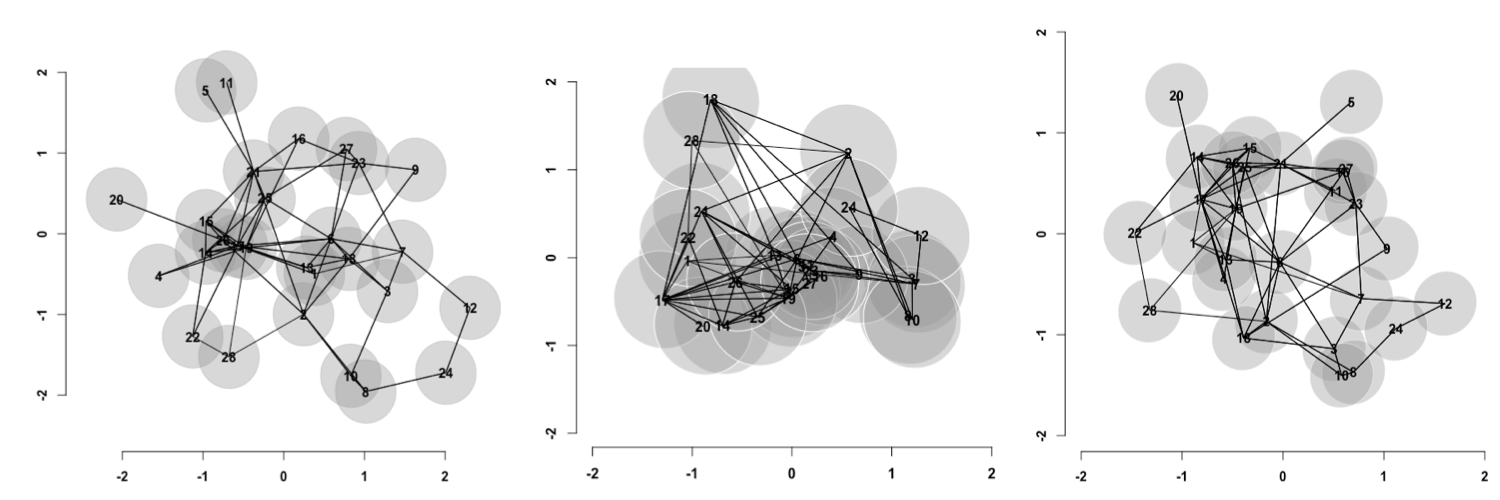}
  \end{center}
  \vspace{-20pt}
      \caption{The estimated latent person positions $\boldsymbol{\tilde{u}_i}$ in the friendship latent space using only the social network $\boldsymbol{Y_I}$ (left), using only the multivariate covariates $\boldsymbol{Y_{IA}}$ (middle) and in the joint latent space using both $\boldsymbol{Y_I}$ and $\boldsymbol{Y_{IA}}$ (right). The edges are the observed friendship connections. The grey ellipses represent $80\%$ approximate credible intervals for the $\tilde{\textbf{u}}_i$. The numbers represent the randomly assigned indices for the French elites. 
      }
\label{persons}
\end{figure}

In Figure \ref{total}, we present the estimated latent person and attribute positions, $\boldsymbol{\tilde{u}_i}$ and $\boldsymbol{\tilde{v}_a}$ from fitting the BLSM (left) and from fitting the APLSM (right). In Figure \ref{persons}, we exclusively compare the resulting latent person positions using only the social network, $\boldsymbol{Y_I}$ (left), only the multivariate covariates, $\boldsymbol{Y_{IA}}$ (middle), and both the social network and the multivariate covariates, $\boldsymbol{Y_I}$ and $\boldsymbol{Y_{IA}}$ (right). An angular rotation of the latent friendship space is applied to match the latent person positions in the joint latent space. A congruence coefficient of $0.96$ was found between the two sets of latent positions. Used as a measure of dimension similarity, a congruence coefficient of $0.95$ and above indicates that the dimensions are identical. A congruence coefficient of $0.91$ was found between the latent person positions using the BLSM and the latent person positions using the APLSM. As expected, the latent person positions' structure using APLSM more closely resembles the structure using LSM than that using BLSM.

\subsubsection*{Assess Model Fit}

 \begin{figure}
\begin{center}
    \includegraphics[width=1\textwidth]{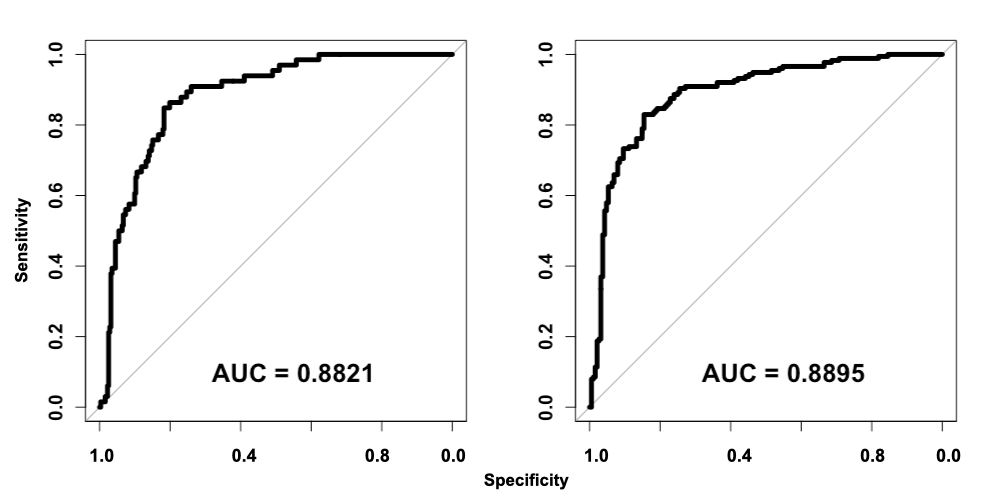}
  \end{center}
  \vspace{-25pt}
      \caption{The ROC curves for $\boldsymbol{Y_I}$ and $\boldsymbol{Y_{IA}}$ using the APLSM}
\label{aucfrench}
\end{figure}

To assess the model fit of APLSM to the data, we obtain the area under the
receiver operating characteristic curve (AUC) of predicting the presence or absence of a link from the estimated link probabilities. The receiver operating characteristic curves (ROCs) and the AUC values for $\boldsymbol{Y_I}$ and $\boldsymbol{Y_{IA}}$ are presented in Figure \ref{aucfrench}. The results show satisfactory fit for both matrices. In addition, we assess whether the APLSM captures the structure of the social network and the dependencies among the attributes and the persons' attributes information in general. 

\begin{figure}
\begin{center}
    \includegraphics[width=1\textwidth]{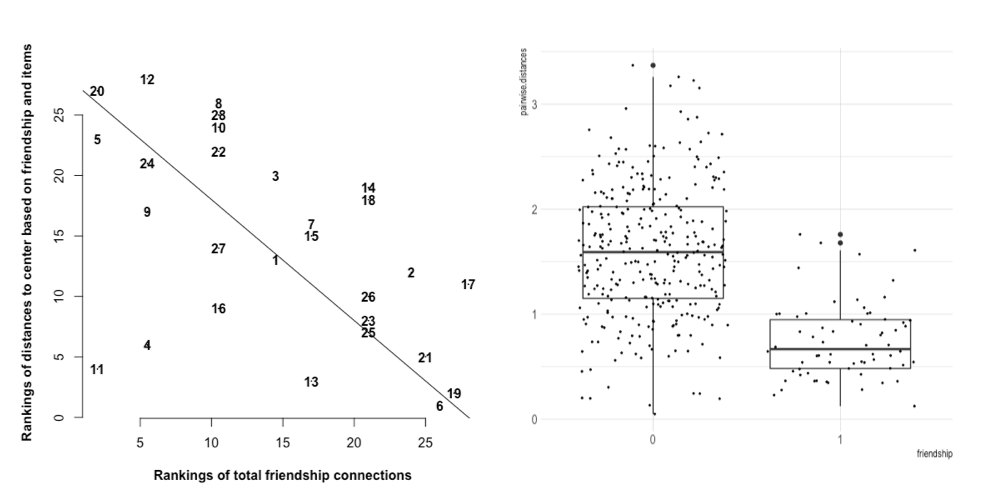}
  \end{center}
  \vspace{-25pt}
      \caption{Assess the fit of the person latent positions. The left panel shows the rankings of nodes' distances to center against the rankings of the nodes' total friendship connections. The solid line is a reference line with an intercept of $N$ and a slope of $-1$. The right panel shows the pairwise distances between pairs of nodes when they are friends versus when they are not friends. }
\label{fitpersons}
\end{figure}

In Figure \ref{fitpersons}, we assess whether the APLSM can capture the friendship structure in the social network. In the left panel, we compare the elite's distances to the center with the sum totals of their friendship connections. We plot the rankings of the distances to center against the rankings of the total friendship connections. In this way, we assess whether the APLSM preserves the elites' social hierarchy without being distracted by the distribution differences between Euclidean distances and friendship counts. A solid reference line is drawn to illustrate the relationship between the two. The intercept for the solid line equals to $N$, the highest ranking, and the slope equals to $-1$.  As the solid line roughly goes through the center of the scatter plot, we know that the rankings of the distances to the center mirror the rankings of the friendship counts in the opposite direction. The Spearman's rank correlation between the Euclidean distances and friendship counts is $-0.53$.  

In the right panel of Figure \ref{fitpersons}, we present the box plot of elites' pairwise differences in the absence of friendship versus when they are friends. We can see that the pairwise distances are generally smaller between elites who are friends with a median of $0.75$, compared to the pairwise distances between elites who are not friends with a median of $1.60$. 
This observation shows that pairwise distances between elites using the APLSM distinguish between whether the elites are friends or not.

\begin{figure}
\begin{center}
    \includegraphics[width=1\textwidth]{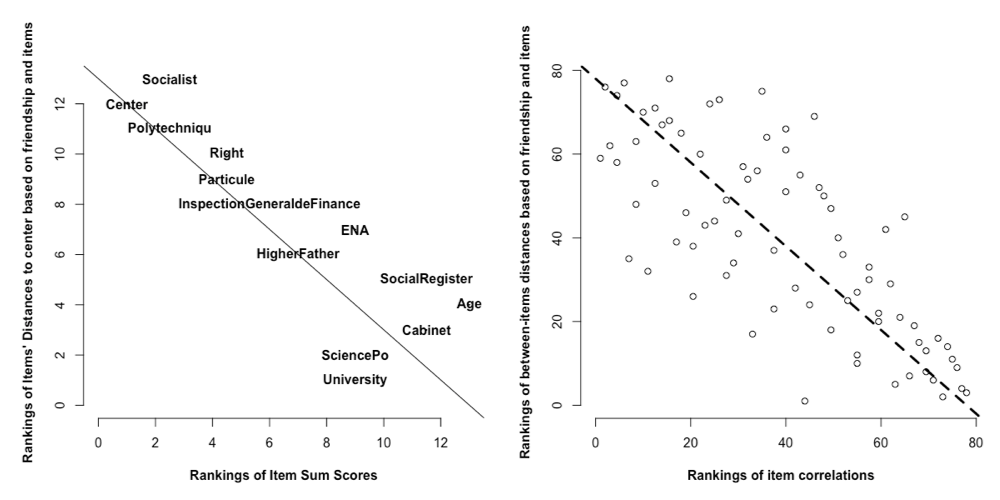}
  \end{center}
  \vspace{-25pt}
      \caption{Assess the fit of the attribute latent positions. The left panel shows the rankings of attributes' distances to center against the rankings of the attributes' sum scores. The right panel shows the rankings of attributes' pairwise distances against the rankings of the attributes' correlations. }
\label{fititems}
\end{figure}

In Figure \ref{fititems}, we assess the APLSM's ability to capture the dependencies among attributes. In the left panel, we plot the rankings of attributes' distances to center against the rankings of the attribute sum scores. As can be seen, the two types of rankings roughly follow the solid line with an intercept of $M$, the highest ranking, and a slope of $-1$, suggesting a mirroring of the two rankings in the opposite direction. The Spearman's rank correlation between the attribute's distances to center and attribute sum scores is $-0.92$. This observation suggests that attributes' distances to the center capture the overall response rates of the attributes. In the right panel of Figure \ref{fititems}, we plot the rankings of the pairwise distances against the rankings of the attribute correlation values. As can be seen, the two types of rankings center around the dashed line with an intercept of $80$, the highest ranking, and a slope of $-1$, suggesting a mirroring of the two rankings in the opposite direction; the Spearman's rank correlation between the two is $-0.62$. This result shows that the pairwise distances between attributes using the APLSM capture the correlations between the attributes.

\begin{figure}[h]
\begin{subfigure}{0.5\textwidth}
  \begin{center}
    \includegraphics[width=\textwidth]{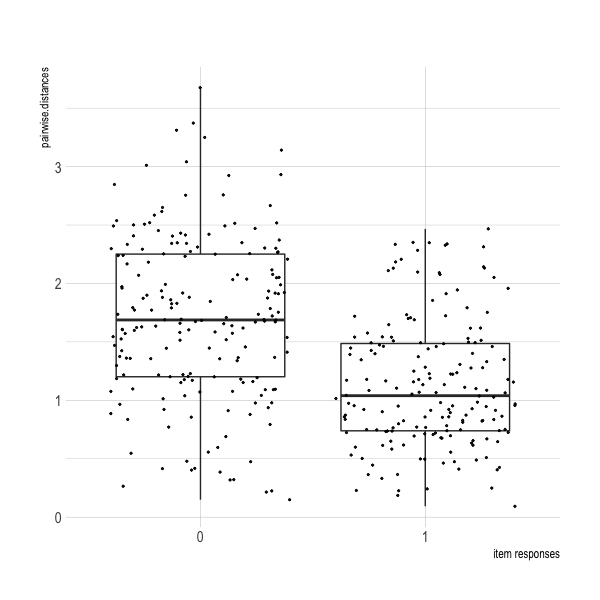}
  \end{center}
  \end{subfigure}%
  \begin{subfigure}{0.5\textwidth}
  \begin{center}
    \includegraphics[width=\textwidth]{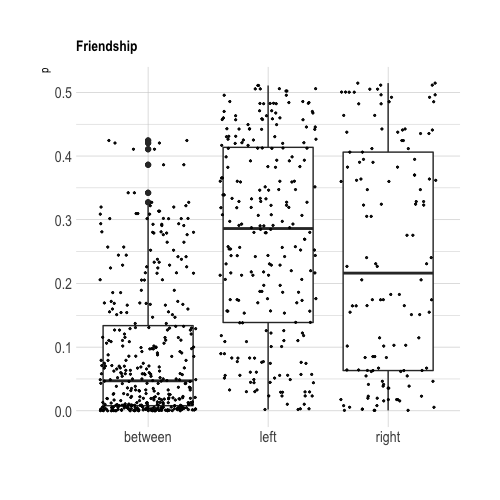}
  \end{center}
  \end{subfigure}
      \caption{(A) The pairwise distances between attributes and persons when the attributes are present versus when the attributes are absent. (B) The box plot showing the probability of a pair of elites being friends in the left and the right clusters and between the two clusters.}
\label{betweenfit}
\end{figure}

Using Figure \ref{betweenfit} (left), we assess whether the APLSM is able to differentiate the presence of an attribute from the absence. We compare the pairwise distances between attributes and persons when the attribute is present to the pairwise distances between persons and attributes when the attribute is absent. We note that when the attributes are present, the distances are generally smaller with a median of $1.13$, than those when the attributes are absent with a median of $1.71$. Therefore the pairwise distances between persons and attributes using the APLSM distinguish the present from the absence of attributes.

\subsubsection*{Redefine the Left and Right Moiety}

 \begin{figure}
\begin{center}
    \includegraphics[width=0.95\textwidth]{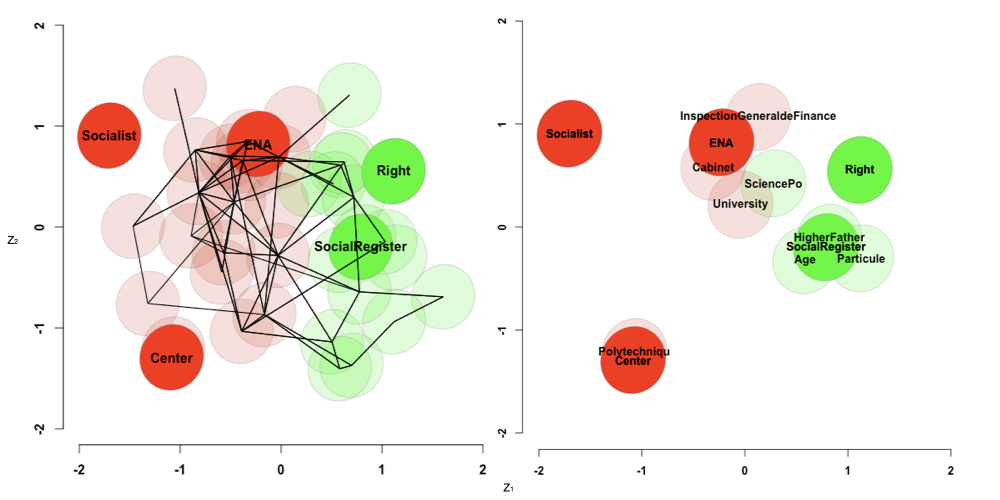}
  \end{center}
  \vspace{-25pt}
      \caption{The division of the French financial elites into two clusters. The left panel displays the two clusters of elites with positions of key attributes. The right panel displays latent attribute positions. Both the latent person and latent attribute positions are part of the joint latent space shown in Figure \ref{total}.  The ellipses represent $80\%$ approximate credible intervals for the latent positions. The solid lines represent the observed friendship connections between pairs of elites.}
\label{2cluster}
\end{figure}

To replicate the left and right moiety in \cite{kadushin1995friendship}, we apply K-means clustering \citep{hartigan1979k} to the latent person and attribute positions with $k=2$. The algorithm partitions the latent positions into $k$ groups which minimize the sum of squares within each group. The k-means clustering is performed with the K-means function in the kknn package, which implements the Hartigan–Wong algorithm \citep{hechenbichler2004weighted}.  We run the K-means algorithm with $100$ random starting positions and take the solution which optimizes the objective function. The resulting two clusters are shown in Figure \ref{2cluster} explaining $50.4\%$ of the total variance in the data (defined as the proportion of the within-cluster sum of squares to the total sum of squares). The two clusters contain both attributes and persons.

In Figure \ref{2cluster}, we show the latent person and attribute positions colored by the resulting two clusters. The solid lines represent the observed friendship connections between pairs of elites. We can see that a more densely connected network can be found on the left side than the right. Figure \ref{betweenfit} (right) shows the probability of a pair of elites being friends in each cluster and between the two clusters. The median probability of friendship connections in the left cluster is roughly $0.29$; the median probability in the right cluster is roughly $0.22$; while between the two clusters, the median probability is roughly $0.05$. This result shows that there is slightly stronger connectivity among elites in the left cluster than in the right cluster. Also, there is a separation of French elites based on social connectivity into the two clusters. 

In Figure \ref{2cluster}, we also see that attributes: Socialists, Inspector General de France, ENA, Cabinet, University, Polytechnique and Center (Centrists) are found in the left (red) cluster, suggesting that the elites in this cluster are more likely to obtain higher education, hold top office positions and identify as socialists or centrists. On the other hand, attributes such as Right (Capitalist), Father Status, Social Register, Particule, Science Po, and Age are found in the right (green) cluster, which suggest that the elites in the right cluster are more likely to be of higher class (being present in the social register, having \textit{particule} in their names and fathers of higher status), older age and identify as politically right.

Two clusters based on APLSM largely correspond to the left and right moieties found in \cite{kadushin1995friendship}. However, in contrast to \cite{kadushin1995friendship}, the two clusters here are joint clusters with both attributes and persons. The addition of the attributes provides more meaning to the division of french elites. We see the division of the social circles based on party affiliations, class, education, and career. This finding was also observed in the previous study though \cite{kadushin1995friendship} compared the background information for elites between the left and right moiety after the analysis of the network (The $12.1 \%$ of the variance in only four attribute variables was explained through regression). Because we are able to use relevant attribute information in the clustering process, there is more confidence in our observation of the division. Besides, the approximate credible intervals of the person and attribute nodes using the APLSM is much smaller than those using the BLSM and the LSM. The drivers behind the division automatically and systematically emerge from the data.

\subsubsection*{Career, Politics and Class}

 \begin{figure}
\begin{center}
    \includegraphics[width=1\textwidth]{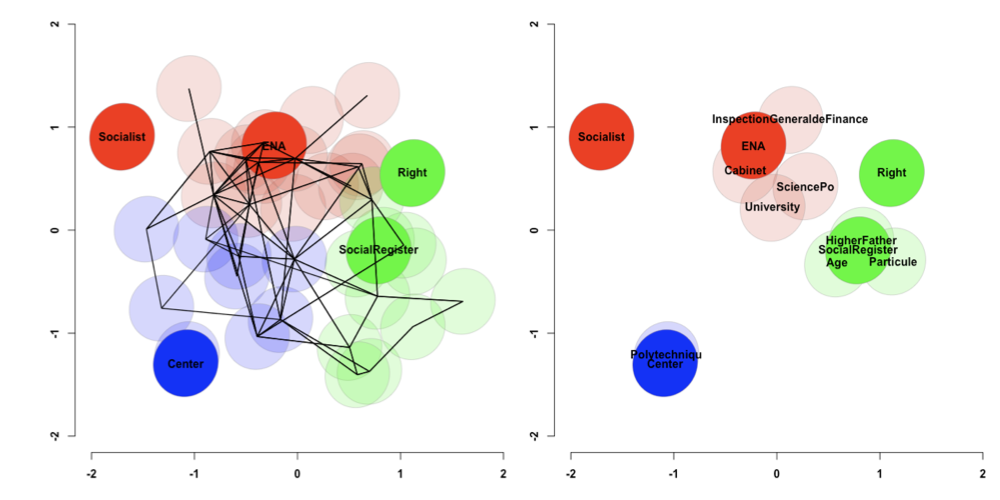}
  \end{center}
  \vspace{-25pt}
      \caption{The division of the French financial elites into three clusters. The left panel displays the three clusters of elites with positions of key attributes. The right panel displays latent attribute positions. Both the latent person and latent attribute positions are part of the joint latent space. The ellipses represent $80\%$ approximate credible intervals for the latent positions. The solid lines represent the observed friendship connections between pairs of elites.}
\label{3cluster}
\end{figure}

The APLSM captures how social relations between French elites and their career, politics, and class information are related. In the previous section, we have replicated the division in the French financial elites between the left and right moiety. Though not mentioned in the previous study
\cite{kadushin1995friendship}, we believe that the presence of a third cluster might be justified given the visible separation between the ENA attribute and the Polytechniqu attribute in the left moiety and our prior knowledge about the French education system. Therefore, we run the k-means clustering algorithm with $k=3$. The algorithm is again implemented with the kmeans function in the kknn P package \citep{hechenbichler2004weighted} with $100$ random starting positions. Approximately $63.9 \%$ of the total variance is explained by the three clusters, with $13.5 \%$ of the variance explained by the added cluster. As before, the resulting three clusters contain both attributes and persons.

Figure \ref{3cluster} displays the estimated latent attribute and person positions (same as before) colored by the resulting three clusters. The additional (blue) cluster is found with the Polytechnique and Center attributes, named the Polytechnique cluster. The red cluster, part of the left moiety, is centered around ENA attributes, called the ENA cluster. The Science Po attribute is now part of the ENA (red) cluster. The green cluster, part of the right moiety, is again centered around high-class attributes, called the HighClass cluster. As we know, Polytechnique, an alternative for the Science Po, indicates an alternative career for the French financial elites.

 \begin{figure}
\begin{center}
    \includegraphics[width=.8\textwidth]{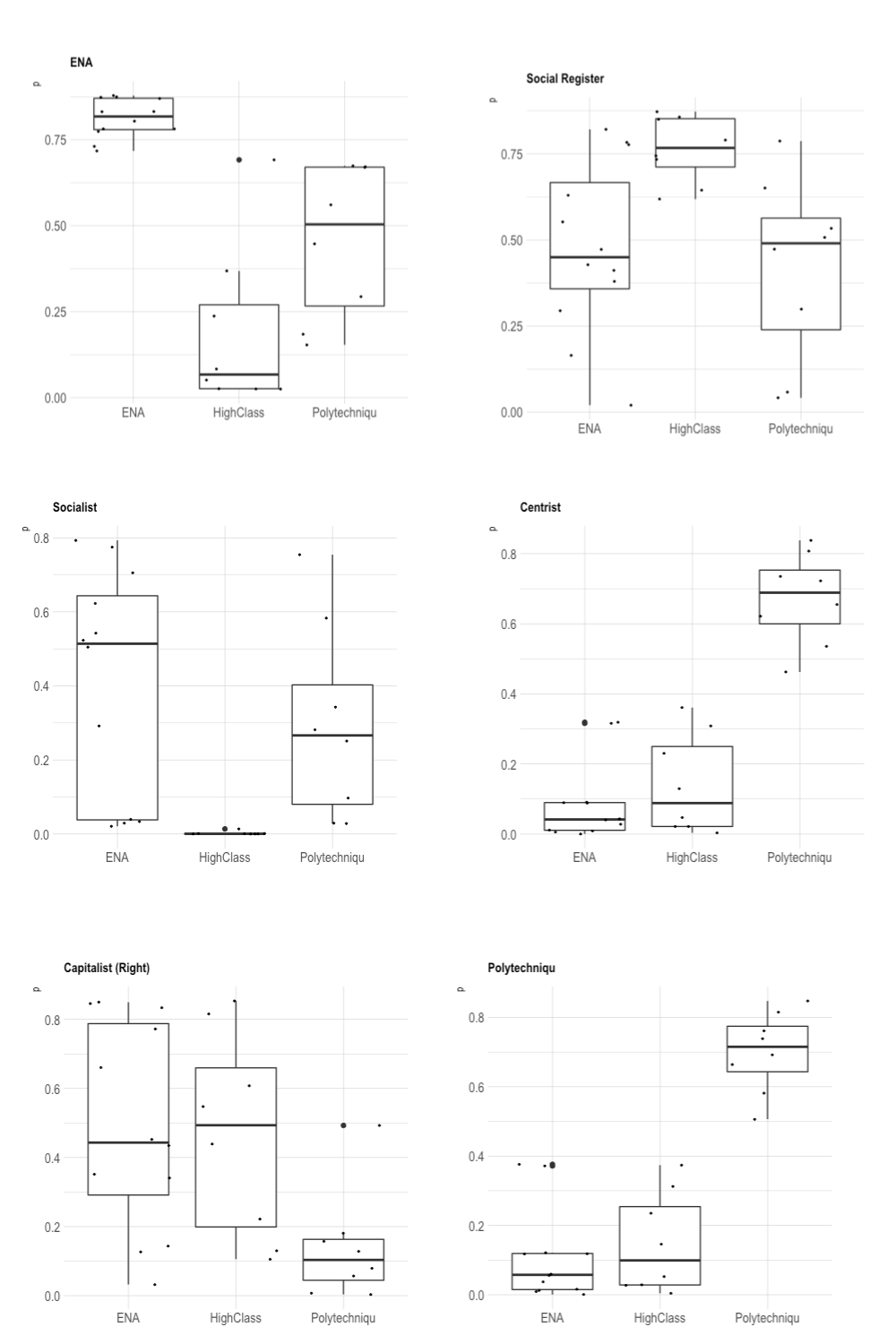}
  \end{center}
  \vspace{-25pt}
      \caption{Boxplots showing the probabilities of feature attributes being indicated as positive in each of the three clusters.}
\label{items3clusters}
\end{figure}

We display the estimated probabilities of the presence of an attribute by elites in each of the three clusters in Figure \ref{items3clusters}. For the ENA attribute, we observe clear distinctions among the three clusters, with the highest probabilities found for elites in the ENA cluster and the lowest probabilities found for elites in the HighClass cluster. For party affiliations (Socialist, Centrist, and Capitalist), we observe higher probabilities of socialists being in the ENA cluster than the Polytechnique cluster and extremely low probabilities of socialists being in the HighClass cluster. We observe slightly higher probabilities of capitalists being in the HighClass cluster than the ENA cluster, though both probabilities are much higher than capitalists being in the Polytechnique cluster. We also observe much higher probabilities of centrists being Polytechnique graduates than being ENA graduates or from a high-class background. For class background and being part of the social register, we observe lower probabilities of being in the social register for elites in the ENA and the Polytechnique clusters than for elites in the HighClass cluster. These additional information can be used to generate new hypotheses for further understanding of the French elite class.

 \begin{figure}
\begin{center}
    \includegraphics[width=.6\textwidth]{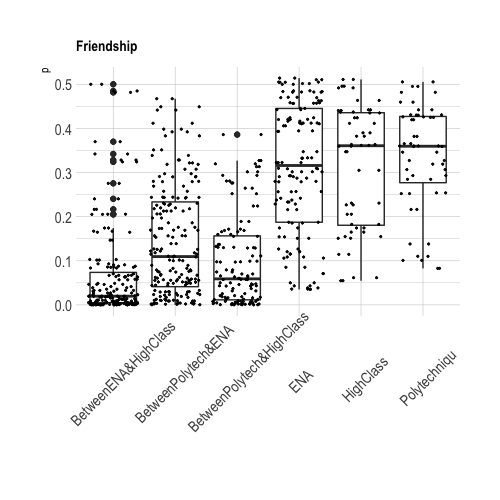}
  \end{center}
  \vspace{-25pt}
      \caption{The boxplot showing the posterior probability of an edge in the social network between elites in the same HighClass, ENA and Polytechniqu clusters and between each pair of clusters}
\label{friend3clusters}
\end{figure}

In Figure \ref{friend3clusters}, we display the probability of a pair of elites being friends with a person from the same cluster and with someone of from each of the other two clusters. The median probabilities of friendship connections within the  HighClass, ENA, and Polytechniqu clusters are $0.36$, $0.32$, and $0.36$ respectively. The median probabilities between clusters are as follows: $0.06$ between the HighClass cluster and the Polytechniqu cluster, $0.02$ between the HighClass cluster and the ENA cluster, and $0.11$ between the ENA cluster and the Polytechniqu cluster. Clearly, there is more intra-cluster connectivity than between-cluster connectivity for all clusters. We observe the strongest connectivity between the elites in the ENA cluster and the Polytechniqu cluster and weakest connectivity between elites in the ENA cluster and the HighClass cluster, which makes sense because one either has the position because of one's education (ENA) or because of the connections of one's family.

 \begin{figure}
\begin{center}
    \includegraphics[width=0.9\textwidth]{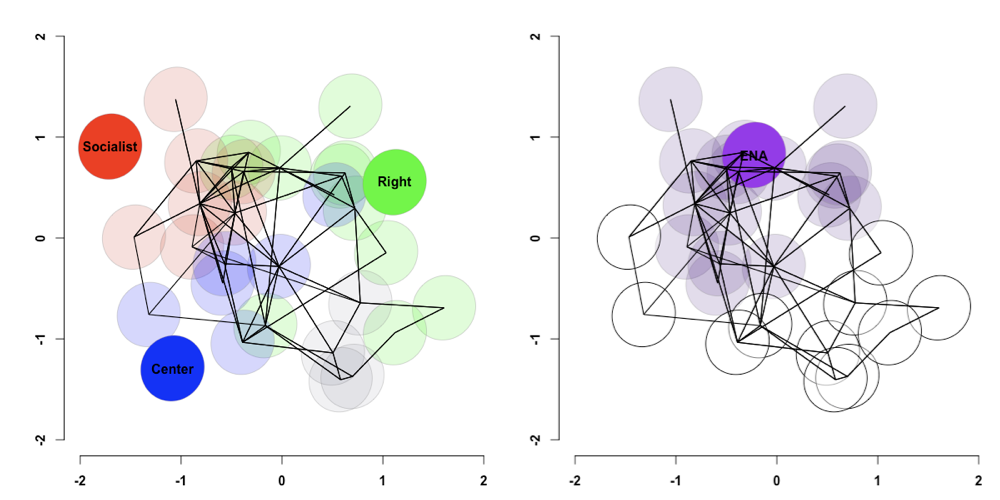}
  \end{center}
  \vspace{-25pt}
      \caption{The latent person positions colored by key attributes: party (left) and ENA (right). The left panel displays the latent positions of socialist (red), centrist (blue) and capitalist (green) elites along with the attributes Socialist, Center and Right. Four French elites indicate no party affiliations and are shown as gray circles in the joint latent space. The right panel displays the latent positions of ENA graduates (purple) along with the attribute ENA. Non-ENA graduates are shown as white circles in the joint latent space. The ellipses represent $80\%$ approximate credible intervals for the latent positions.}
\label{personsIV}
\end{figure}

The left panel of Figure \ref{personsIV} displays the latent person positions of socialists (red), centrists (blue), and capitalists (green) elites along with the latent attribute position of attributes Socialist, Center, and Right. Elites with no party affiliations are shown as gray circles in the joint latent space. As can be seen, elites with different party affiliations are positioned near the corresponding attributes and are far apart in the joint latent space. The right panel of Figure \ref{personsIV} displays the latent positions of ENA graduates (purple) and non-ENA graduates (white) along with the attribute ENA. Again, we see a clear separation between ENA graduates and non-ENA graduates as the ENA graduates center around the ENA attribute.

\section{Discussions and conclusion}

The APLSM outlined in this article constitutes a principle strategy for jointly analyzing social networks and high dimensional multivariate covariates. We have argued for and presented evidence that a joint analysis of friendships and individual outcomes is crucial in understanding human behaviors. In particular, using APLSM, we have analyzed the French financial elite data, replicated the division in the elite circle between the left and right moiety, explained and consolidated the division using career, class, and political information. Furthermore, we identified an additional cluster in the joint latent space. Therefore, we believe that APLSM can significantly help researchers understand how friendships and multivariate covariates are intertwined and inspire further model development in this area.  

\section*{Supplementary Materials}

The supplementary materials contain detailed derivations of the variational bayesian EM algorithm for the proposed model, the APLSM, parts of which are used to estimate the BLSM.

\medskip

\bibliography{cluster,vc,hetgen,triads,triads1,GraphTheory,Networks,psychology,references,irt,cand,coevol,sarpeer}

\begin{thebibliography}{87}
\newcommand{\enquote}[1]{``#1''}
\expandafter\ifx\csname natexlab\endcsname\relax\def\natexlab#1{#1}\fi

\bibitem[{Agarwal and Xue(2020)}]{agarwal2019model}
Agarwal, A.,  and Xue, L. (2020), \enquote{Model-based clustering of
  nonparametric weighted networks with application to water pollution
  analysis,} \textit{Technometrics}, 62, 161--172.

\bibitem[{Airoldi et~al.(2008)Airoldi, Blei, Fienberg, and
  Xing}]{airoldi2008mixed}
Airoldi, E.~M., Blei, D.~M., Fienberg, S.~E.,  and Xing, E.~P. (2008),
  \enquote{Mixed membership stochastic blockmodels,} \textit{Journal of Machine
  Learning Research}, 9, 1981--2014.

\bibitem[{Albert and Barab{\'a}si(2002)}]{albert2002statistical}
Albert, R.,  and Barab{\'a}si, A.-L. (2002), \enquote{Statistical mechanics of
  complex networks,} \textit{Reviews of modern physics}, 74, 47.

\bibitem[{Arroyo et~al.(2019)Arroyo, Athreya, Cape, Chen, Priebe, and
  Vogelstein}]{arroyo2019inference}
Arroyo, J., Athreya, A., Cape, J., Chen, G., Priebe, C.~E.,  and Vogelstein,
  J.~T. (2019), \enquote{Inference for multiple heterogeneous networks with a
  common invariant subspace,} \textit{arXiv preprint arXiv:1906.10026}.

\bibitem[{Attias(1999)}]{attias1999inferring}
Attias, H. (1999), \enquote{Inferring parameters and structure of latent
  variable models by variational Bayes,} in \textit{Proceedings of the
  Fifteenth conference on Uncertainty in artificial intelligence}, Morgan
  Kaufmann Publishers Inc., pp. 21--30.

\bibitem[{Austin et~al.(2013)Austin, Linkletter, and Wu}]{austin2013covariate}
Austin, A., Linkletter, C.,  and Wu, Z. (2013), \enquote{Covariate-defined
  latent space random effects model,} \textit{Social networks}, 35, 338--346.

\bibitem[{Barab{\'a}si and Albert(1999)}]{barabasi1999emergence}
Barab{\'a}si, A.-L.,  and Albert, R. (1999), \enquote{Emergence of scaling in
  random networks,} \textit{science}, 286, 509--512.

\bibitem[{Barbillon et~al.(2015)Barbillon, Donnet, Lazega, and
  Bar-Hen}]{barbillon2015stochastic}
Barbillon, P., Donnet, S., Lazega, E.,  and Bar-Hen, A. (2015),
  \enquote{Stochastic block models for multiplex networks: an application to
  networks of researchers,} \textit{arXiv preprint arXiv:1501.06444}.

\bibitem[{Baum et~al.(1970)Baum, Petrie, Soules, and
  Weiss}]{baum1970maximization}
Baum, L.~E., Petrie, T., Soules, G.,  and Weiss, N. (1970), \enquote{A
  maximization technique occurring in the statistical analysis of probabilistic
  functions of Markov chains,} \textit{The annals of mathematical statistics},
  41, 164--171.

\bibitem[{Beal et~al.(2006)Beal, Ghahramani, et~al.}]{beal2006variational}
Beal, M.~J., Ghahramani, Z. et~al. (2006), \enquote{Variational Bayesian
  learning of directed graphical models with hidden variables,}
  \textit{Bayesian Analysis}, 1, 793--831.

\bibitem[{Beal et~al.(2003)}]{beal2003variational}
Beal, M.~J. et~al. (2003), \textit{Variational algorithms for approximate
  Bayesian inference}, university of London London.

\bibitem[{Bickel and Chen(2009)}]{bc09}
Bickel, P.~J.,  and Chen, A. (2009), \enquote{A nonparametric view of network
  models and Newman--Girvan and other modularities,} \textit{Proceedings of the
  National Academy of Sciences}, 106, 21068--21073.

\bibitem[{Blei et~al.(2017)Blei, Kucukelbir, and
  McAuliffe}]{blei2017variational}
Blei, D.~M., Kucukelbir, A.,  and McAuliffe, J.~D. (2017), \enquote{Variational
  inference: A review for statisticians,} \textit{Journal of the American
  Statistical Association}, 112, 859--877.

\bibitem[{Boccaletti et~al.(2014)Boccaletti, Bianconi, Criado, Del~Genio,
  G{\'o}mez-Garde{\~n}es, Romance, Sendina-Nadal, Wang, and
  Zanin}]{boccaletti14}
Boccaletti, S., Bianconi, G., Criado, R., Del~Genio, C.~I.,
  G{\'o}mez-Garde{\~n}es, J., Romance, M., Sendina-Nadal, I., Wang, Z.,  and
  Zanin, M. (2014), \enquote{The structure and dynamics of multilayer
  networks,} \textit{Physics Reports}, 544, 1--122.

\bibitem[{Borgatti et~al.(2009)Borgatti, Mehra, Brass, and
  Labianca}]{borgatti2009network}
Borgatti, S.~P., Mehra, A., Brass, D.~J.,  and Labianca, G. (2009),
  \enquote{Network analysis in the social sciences,} \textit{science}, 323,
  892--895.

\bibitem[{Bramoull{\'e} et~al.(2009)Bramoull{\'e}, Djebbari, and
  Fortin}]{bramoulle2009identification}
Bramoull{\'e}, Y., Djebbari, H.,  and Fortin, B. (2009),
  \enquote{Identification of peer effects through social networks,}
  \textit{Journal of econometrics}, 150, 41--55.

\bibitem[{Bullmore and Sporns(2009)}]{bullmore09}
Bullmore, E.,  and Sporns, O. (2009), \enquote{Complex brain networks: graph
  theoretical analysis of structural and functional systems,} \textit{Nature
  Reviews Neuroscience}, 10, 186--198.

\bibitem[{Carrington et~al.(2005)Carrington, Scott, and
  Wasserman}]{carrington2005models}
Carrington, P.~J., Scott, J.,  and Wasserman, S. (2005), \textit{Models and
  methods in social network analysis}, vol.~28, Cambridge university press.

\bibitem[{Celisse et~al.(2012)Celisse, Daudin, and Pierre}]{cdp11}
Celisse, A., Daudin, J.~J.,  and Pierre, L. (2012), \enquote{Consistency of
  maximum-likelihood and variational estimators in the stochastic block model,}
  \textit{Electronic Journal of Statistics}, 6, 1847--1899.

\bibitem[{D'Angelo et~al.(2018)D'Angelo, Alf{\`o}, and Murphy}]{d2018node}
D'Angelo, S., Alf{\`o}, M.,  and Murphy, T.~B. (2018), \enquote{Node-specific
  effects in latent space modelling of multidimensional networks,} in
  \textit{49th Scientific meeting of the Italian Statistical Society}.

\bibitem[{Daudin et~al.(2008)Daudin, Picard, and Robin}]{dpr08}
Daudin, J.~J., Picard, F.,  and Robin, S. (2008), \enquote{A mixture model for
  random graphs,} \textit{Stat Comput}, 18, 173--183.

\bibitem[{Dean et~al.(2017)Dean, Bauer, and Prinstein}]{dean2017friendship}
Dean, D.~O., Bauer, D.~J.,  and Prinstein, M.~J. (2017), \enquote{Friendship
  Dissolution Within Social Networks Modeled Through Multilevel Event History
  Analysis,} \textit{Multivariate behavioral research}, 52, 271--289.

\bibitem[{Dempster et~al.(1977)Dempster, Laird, and
  Rubin}]{dempster1977maximum}
Dempster, A.~P., Laird, N.~M.,  and Rubin, D.~B. (1977), \enquote{Maximum
  likelihood from incomplete data via the EM algorithm,} \textit{Journal of the
  Royal Statistical Society. Series B (Methodological)}, 1--38.

\bibitem[{Dorans and Drasgow(1978)}]{dorans1978alternative}
Dorans, N.,  and Drasgow, F. (1978), \enquote{Alternative weighting schemes for
  linear prediction,} \textit{Organizational Behavior and Human Performance},
  21, 316--345.

\bibitem[{Ferrara et~al.(2014)Ferrara, Interdonato, and
  Tagarelli}]{ferrara2014online}
Ferrara, E., Interdonato, R.,  and Tagarelli, A. (2014), \enquote{Online
  popularity and topical interests through the lens of instagram,} in
  \textit{Proceedings of the 25th ACM conference on Hypertext and social
  media}, ACM, pp. 24--34.

\bibitem[{Fosdick and Hoff(2015)}]{fosdick2015testing}
Fosdick, B.~K.,  and Hoff, P.~D. (2015), \enquote{Testing and modeling
  dependencies between a network and nodal attributes,} \textit{Journal of the
  American Statistical Association}, 110, 1047--1056.

\bibitem[{Frank et~al.(2004)Frank, Zhao, and Borman}]{frank2004social}
Frank, K.~A., Zhao, Y.,  and Borman, K. (2004), \enquote{Social capital and the
  diffusion of innovations within organizations: The case of computer
  technology in schools,} \textit{Sociology of education}, 77, 148--171.

\bibitem[{Fratiglioni et~al.(2000)Fratiglioni, Wang, Ericsson, Maytan, and
  Winblad}]{fratiglioni2000influence}
Fratiglioni, L., Wang, H.-X., Ericsson, K., Maytan, M.,  and Winblad, B.
  (2000), \enquote{Influence of social network on occurrence of dementia: a
  community-based longitudinal study,} \textit{The lancet}, 355, 1315--1319.

\bibitem[{Friel et~al.(2016)Friel, Rastelli, Wyse, and
  Raftery}]{friel2016interlocking}
Friel, N., Rastelli, R., Wyse, J.,  and Raftery, A.~E. (2016),
  \enquote{Interlocking directorates in Irish companies using a latent space
  model for bipartite networks,} \textit{Proceedings of the National Academy of
  Sciences}, 113, 6629--6634.

\bibitem[{Fujimoto et~al.(2013)Fujimoto, Wang, and
  Valente}]{fujimoto2013decomposed}
Fujimoto, K., Wang, P.,  and Valente, T.~W. (2013), \enquote{The decomposed
  affiliation exposure model: A network approach to segregating peer influences
  from crowds and organized sports,} \textit{Network Science}, 1, 154--169.

\bibitem[{Girvan and Newman(2002)}]{girvan2002community}
Girvan, M.,  and Newman, M.~E. (2002), \enquote{Community structure in social
  and biological networks,} \textit{Proceedings of the national academy of
  sciences}, 99, 7821--7826.

\bibitem[{Goldsmith-Pinkham and Imbens(2013)}]{goldsmith2013social}
Goldsmith-Pinkham, P.,  and Imbens, G.~W. (2013), \enquote{Social networks and
  the identification of peer effects,} \textit{Journal of Business \& Economic
  Statistics}, 31, 253--264.

\bibitem[{Gollini and Murphy(2016)}]{gollini2016joint}
Gollini, I.,  and Murphy, T.~B. (2016), \enquote{Joint modeling of multiple
  network views,} \textit{Journal of Computational and Graphical Statistics},
  25, 246--265.

\bibitem[{Guhaniyogi et~al.(2020)Guhaniyogi, Rodriguez,
  et~al.}]{guhaniyogi2020joint}
Guhaniyogi, R., Rodriguez, A. et~al. (2020), \enquote{Joint modeling of
  longitudinal relational data and exogenous variables,} \textit{Bayesian
  Analysis}, 15, 477--503.

\bibitem[{Handcock et~al.(2007)Handcock, Raftery, and Tantrum}]{hrt07}
Handcock, M.~S., Raftery, A.~E.,  and Tantrum, J.~M. (2007),
  \enquote{Model-based clustering for social networks,} \textit{J. Roy.
  Statist. Soc. Ser. A}, 170, 301--354.

\bibitem[{Hartigan et~al.(1979)Hartigan, Wong, et~al.}]{hartigan1979k}
Hartigan, J., Wong, M. et~al. (1979), \enquote{A k-means clustering algorithm,}
  \textit{New Haven}.

\bibitem[{He et~al.(2014)He, Kan, Xie, and Chen}]{he2014comment}
He, X., Kan, M.-Y., Xie, P.,  and Chen, X. (2014), \enquote{Comment-based
  multi-view clustering of web 2.0 items,} in \textit{Proceedings of the 23rd
  international conference on World wide web}, ACM, pp. 771--782.

\bibitem[{Hechenbichler and Schliep(2004)}]{hechenbichler2004weighted}
Hechenbichler, K.,  and Schliep, K. (2004), \enquote{Weighted
  k-nearest-neighbor techniques and ordinal classification,} .

\bibitem[{Henderson and Searle(1981)}]{henderson1981deriving}
Henderson, H.~V.,  and Searle, S.~R. (1981), \enquote{On deriving the inverse
  of a sum of matrices,} \textit{Siam Review}, 23, 53--60.

\bibitem[{Hoff(2008)}]{hoff2008modeling}
Hoff, P. (2008), \enquote{Modeling homophily and stochastic equivalence in
  symmetric relational data,} in \textit{Advances in neural information
  processing systems}, pp. 657--664.

\bibitem[{Hoff(2005)}]{hoff2005bilinear}
Hoff, P.~D. (2005), \enquote{Bilinear mixed-effects models for dyadic data,}
  \textit{Journal of the american Statistical association}, 100, 286--295.

\bibitem[{Hoff(2009)}]{hoff2009multiplicative}
--- (2009), \enquote{Multiplicative latent factor models for description and
  prediction of social networks,} \textit{Computational and mathematical
  organization theory}, 15, 261.

\bibitem[{Hoff(2018)}]{hoff2018additive}
--- (2018), \enquote{Additive and multiplicative effects network models,}
  \textit{arXiv preprint arXiv:1807.08038}.

\bibitem[{Hoff et~al.(2002)Hoff, Raftery, and Handcock}]{hoff2002latent}
Hoff, P.~D., Raftery, A.~E.,  and Handcock, M.~S. (2002), \enquote{Latent space
  approaches to social network analysis,} \textit{Journal of the american
  Statistical association}, 97, 1090--1098.

\bibitem[{Huang et~al.(2020)Huang, Liu, Chen, et~al.}]{huang2020mixed}
Huang, W., Liu, Y., Chen, Y. et~al. (2020), \enquote{Mixed membership
  stochastic blockmodels for heterogeneous networks,} \textit{Bayesian
  Analysis}, 15, 711--736.

\bibitem[{Jackson et~al.(2008)}]{jackson2008social}
Jackson, M.~O. et~al. (2008), \textit{Social and economic networks}, vol.~3,
  Princeton University Press Princeton.

\bibitem[{Jensen(1906)}]{jensen1906fonctions}
Jensen, J. L. W.~V. (1906), \enquote{Sur les fonctions convexes et les
  in{\'e}galit{\'e}s entre les valeurs moyennes,} \textit{Acta mathematica},
  30, 175--193.

\bibitem[{Jordan et~al.(1999)Jordan, Ghahramani, Jaakkola, and
  Saul}]{jordan1999introduction}
Jordan, M.~I., Ghahramani, Z., Jaakkola, T.~S.,  and Saul, L.~K. (1999),
  \enquote{An introduction to variational methods for graphical models,}
  \textit{Machine learning}, 37, 183--233.

\bibitem[{Kadushin(1995)}]{kadushin1995friendship}
Kadushin, C. (1995), \enquote{Friendship among the French financial elite,}
  \textit{American Sociological Review}, 202--221.

\bibitem[{K{\'e}fi et~al.(2016)K{\'e}fi, Miele, Wieters, Navarrete, and
  Berlow}]{kefi2016structured}
K{\'e}fi, S., Miele, V., Wieters, E.~A., Navarrete, S.~A.,  and Berlow, E.~L.
  (2016), \enquote{How structured is the entangled bank? The surprisingly
  simple organization of multiplex ecological networks leads to increased
  persistence and resilience,} \textit{PLoS biology}, 14, e1002527.

\bibitem[{Kim and Srivastava(2007)}]{kim2007impact}
Kim, Y.,  and Srivastava, J. (2007), \enquote{Impact of social influence in
  e-commerce decision making,} in \textit{Proceedings of the ninth
  international conference on Electronic commerce}, ACM, pp. 293--302.

\bibitem[{Kivel{\"a} et~al.(2014)Kivel{\"a}, Arenas, Barthelemy, Gleeson,
  Moreno, and Porter}]{kivela14}
Kivel{\"a}, M., Arenas, A., Barthelemy, M., Gleeson, J.~P., Moreno, Y.,  and
  Porter, M.~A. (2014), \enquote{Multilayer networks,} \textit{Journal of
  Complex Networks}, 2, 203--271.

\bibitem[{Krivitsky and Handcock(2008)}]{krivitsky2008fitting}
Krivitsky, P.~N.,  and Handcock, M.~S. (2008), \enquote{Fitting position latent
  cluster models for social networks with latentnet,} \textit{Journal of
  Statistical Software}, 24.

\bibitem[{Krivitsky et~al.(2009)Krivitsky, Handcock, Raftery, and
  Hoff}]{krivitsky2009representing}
Krivitsky, P.~N., Handcock, M.~S., Raftery, A.~E.,  and Hoff, P.~D. (2009),
  \enquote{Representing degree distributions, clustering, and homophily in
  social networks with latent cluster random effects models,} \textit{Social
  networks}, 31, 204--213.

\bibitem[{Kwon et~al.(2014)Kwon, Stefanone, and Barnett}]{kwon2014social}
Kwon, K.~H., Stefanone, M.~A.,  and Barnett, G.~A. (2014), \enquote{Social
  network influence on online behavioral choices: exploring group formation on
  social network sites,} \textit{American Behavioral Scientist}, 58,
  1345--1360.

\bibitem[{Lazer(2011)}]{lazer2011networks}
Lazer, D. (2011), \enquote{Networks in political science: Back to the future,}
  \textit{PS: Political Science \& Politics}, 44, 61--68.

\bibitem[{Leenders(2002)}]{leenders2002modeling}
Leenders, R. T.~A. (2002), \enquote{Modeling social influence through network
  autocorrelation: constructing the weight matrix,} \textit{Social networks},
  24, 21--47.

\bibitem[{Liu et~al.(2014)Liu, Liu, Murata, and Wakita}]{lmw14}
Liu, X., Liu, W., Murata, T.,  and Wakita, K. (2014), \enquote{A framework for
  community detection in heterogeneous multi-relational networks,}
  \textit{Advances in Complex Systems}, 17, 1450018.

\bibitem[{Lusher et~al.(2013)Lusher, Koskinen, and
  Robins}]{lusher2013exponential}
Lusher, D., Koskinen, J.,  and Robins, G. (2013), \textit{Exponential random
  graph models for social networks: Theory, methods, and applications},
  Cambridge University Press.

\bibitem[{Ma et~al.(2020)Ma, Ma, and Yuan}]{ma2020universal}
Ma, Z., Ma, Z.,  and Yuan, H. (2020), \enquote{Universal Latent Space Model
  Fitting for Large Networks with Edge Covariates.} \textit{Journal of Machine
  Learning Research}, 21, 1--67.

\bibitem[{Matias and Miele(2016)}]{matias15}
Matias, C.,  and Miele, V. (2016), \enquote{Statistical clustering of temporal
  networks through a dynamic stochastic block model,} \textit{Journal of the
  Royal Statistical Society: Series B (Statistical Methodology)}.

\bibitem[{McPherson et~al.(2001)McPherson, Smith-Lovin, and
  Cook}]{mcpherson2001birds}
McPherson, M., Smith-Lovin, L.,  and Cook, J.~M. (2001), \enquote{Birds of a
  feather: Homophily in social networks,} \textit{Annual review of sociology},
  27, 415--444.

\bibitem[{Mele et~al.(2019)Mele, Hao, Cape, and Priebe}]{mele2019spectral}
Mele, A., Hao, L., Cape, J.,  and Priebe, C.~E. (2019), \enquote{Spectral
  inference for large Stochastic Blockmodels with nodal covariates,}
  \textit{arXiv preprint arXiv:1908.06438}.

\bibitem[{Mercken et~al.(2010)Mercken, Snijders, Steglich, Vartiainen, and
  De~Vries}]{mercken2010dynamics}
Mercken, L., Snijders, T.~A., Steglich, C., Vartiainen, E.,  and De~Vries, H.
  (2010), \enquote{Dynamics of adolescent friendship networks and smoking
  behavior,} \textit{Social networks}, 32, 72--81.

\bibitem[{Mucha et~al.(2010)Mucha, Richardson, Macon, Porter, and
  Onnela}]{mucha10}
Mucha, P.~J., Richardson, T., Macon, K., Porter, M.~A.,  and Onnela, J.~P.
  (2010), \enquote{Community structure in time-dependent, multiscale, and
  multiplex networks,} \textit{Science}, 328, 876--878.

\bibitem[{Nickel et~al.(2016)Nickel, Murphy, Tresp, and
  Gabrilovich}]{nickel2016review}
Nickel, M., Murphy, K., Tresp, V.,  and Gabrilovich, E. (2016), \enquote{A
  review of relational machine learning for knowledge graphs,}
  \textit{Proceedings of the IEEE}, 104, 11--33.

\bibitem[{Paul and Chen(2016)}]{pc15}
Paul, S.,  and Chen, Y. (2016), \enquote{Consistent community detection in
  multi-relational data through restricted multi-layer stochastic blockmodel,}
  \textit{Electronic Journal of Statistics}, 10, 3807--3870.

\bibitem[{Paul et~al.(2020{\natexlab{a}})Paul, Chen, et~al.}]{paul2020random}
Paul, S., Chen, Y. et~al. (2020{\natexlab{a}}), \enquote{A random effects
  stochastic block model for joint community detection in multiple networks
  with applications to neuroimaging,} \textit{Annals of Applied Statistics},
  14, 993--1029.

\bibitem[{Paul et~al.(2020{\natexlab{b}})Paul, Chen, et~al.}]{paul2020spectral}
--- (2020{\natexlab{b}}), \enquote{Spectral and matrix factorization methods
  for consistent community detection in multi-layer networks,} \textit{The
  Annals of Statistics}, 48, 230--250.

\bibitem[{Robins et~al.(2001)Robins, Pattison, and Elliott}]{robins2001network}
Robins, G., Pattison, P.,  and Elliott, P. (2001), \enquote{Network models for
  social influence processes,} \textit{Psychometrika}, 66, 161--189.

\bibitem[{Rubinov and Sporns(2010)}]{rubinov10}
Rubinov, M.,  and Sporns, O. (2010), \enquote{Complex network measures of brain
  connectivity: uses and interpretations,} \textit{Neuroimage}, 52, 1059--1069.

\bibitem[{Salter-Townshend and McCormick(2017)}]{salter2017latent}
Salter-Townshend, M.,  and McCormick, T.~H. (2017), \enquote{Latent space
  models for multiview network data,} \textit{The annals of applied
  statistics}, 11, 1217.

\bibitem[{Salter-Townshend and Murphy(2013)}]{salter2013variational}
Salter-Townshend, M.,  and Murphy, T.~B. (2013), \enquote{Variational Bayesian
  inference for the latent position cluster model for network data,}
  \textit{Computational Statistics \& Data Analysis}, 57, 661--671.

\bibitem[{Sengupta and Chen(2015)}]{sengupta2015spectral}
Sengupta, S.,  and Chen, Y. (2015), \enquote{Spectral clustering in
  heterogeneous networks,} \textit{Statistica Sinica}, 1081--1106.

\bibitem[{Sewell and Chen(2015)}]{sewell2015latent}
Sewell, D.~K.--- (2015), \enquote{Latent space models for dynamic networks,}
  \textit{Journal of the American Statistical Association}, 110, 1646--1657.

\bibitem[{Sewell and Chen(2016)}]{sewell2016latent}
Sewell, D.~K.--- (2016), \enquote{Latent space models for dynamic networks with
  weighted edges,} \textit{Social Networks}, 44, 105--116.

\bibitem[{Shalizi and Thomas(2011)}]{shalizi2011homophily}
Shalizi, C.~R.,  and Thomas, A.~C. (2011), \enquote{Homophily and contagion are
  generically confounded in observational social network studies,}
  \textit{Sociological methods \& research}, 40, 211--239.

\bibitem[{Shmulevich et~al.(2002)Shmulevich, Dougherty, Kim, and
  Zhang}]{shmulevich2002probabilistic}
Shmulevich, I., Dougherty, E.~R., Kim, S.,  and Zhang, W. (2002),
  \enquote{Probabilistic Boolean networks: a rule-based uncertainty model for
  gene regulatory networks,} \textit{Bioinformatics}, 18, 261--274.

\bibitem[{Sun et~al.(2009)Sun, Yu, and Han}]{sun2009ranking}
Sun, Y., Yu, Y.,  and Han, J. (2009), \enquote{Ranking-based clustering of
  heterogeneous information networks with star network schema,} in
  \textit{Proceedings of the 15th ACM SIGKDD international conference on
  Knowledge discovery and data mining}, ACM, pp. 797--806.

\bibitem[{Sweet and Adhikari(2020)}]{sweet2020latent}
Sweet, T.,  and Adhikari, S. (2020), \enquote{A Latent Space Network Model for
  Social Influence,} \textit{Psychometrika}, 1--24.

\bibitem[{Sweet(2015)}]{sweet2015incorporating}
Sweet, T.~M. (2015), \enquote{Incorporating covariates into stochastic
  blockmodels,} \textit{Journal of Educational and Behavioral Statistics}, 40,
  635--664.

\bibitem[{VanderWeele(2011)}]{vanderweele2011sensitivity}
VanderWeele, T.~J. (2011), \enquote{Sensitivity analysis for contagion effects
  in social networks,} \textit{Sociological Methods \& Research}, 40, 240--255.

\bibitem[{VanderWeele and An(2013)}]{vanderweele2013social}
VanderWeele, T.~J.,  and An, W. (2013), \enquote{Social networks and causal
  inference,} in \textit{Handbook of causal analysis for social research},
  Springer, pp. 353--374.

\bibitem[{Wasserman et~al.(1994)Wasserman, Faust, et~al.}]{wasserman1994social}
Wasserman, S., Faust, K. et~al. (1994), \textit{Social network analysis:
  Methods and applications}, vol.~8, Cambridge university press.

\bibitem[{Watts and Strogatz(1998)}]{watts1998collective}
Watts, D.~J.,  and Strogatz, S.~H. (1998), \enquote{Collective dynamics of
  ‘small-world’networks,} \textit{nature}, 393, 440.

\bibitem[{Xu et~al.(2014)Xu, Kliger, and Hero~Iii}]{xu14}
Xu, K.~S., Kliger, M.,  and Hero~Iii, A.~O. (2014), \enquote{Adaptive
  evolutionary clustering,} \textit{Data Mining and Knowledge Discovery}, 28,
  304--336.

\bibitem[{Zhang et~al.(2020)Zhang, Xue, and Zhu}]{zhang2020flexible}
Zhang, X., Xue, S.,  and Zhu, J. (2020), \enquote{A Flexible Latent Space Model
  for Multilayer Networks,} in \textit{International Conference on Machine
  Learning}, PMLR, pp. 11288--11297.

\end{thebibliography}

\newpage

{\large\bf Supplementary Material for "Joint Latent Space Model for Social Networks with Multivariate Attributes".}

\section{The Estimation Procedure for APLSM}
\subsection{Derivation of KL Divergence}
We set the variational parameter as $\Theta = \tilde{\alpha}_0,\tilde{\alpha}_1$ and $\boldsymbol{\tilde{u}_i},\tilde{\Lambda}_{0},\boldsymbol{\tilde{v}_a},\tilde{\Lambda}_{1}$, where $q(\boldsymbol{u_i})=N(\boldsymbol{\tilde{u}_i},\tilde{\Lambda}_{0})$, and $q(\boldsymbol{v_a})=N(\boldsymbol{\tilde{v}_a},\tilde{\Lambda}_{1})$. We set the variational posterior as:
\begin{equation*}
q(\boldsymbol{U},\boldsymbol{V} |\boldsymbol{Y_I}, \boldsymbol{Y_{IA}})= \prod_{i=1}^Nq(\boldsymbol{u_i}) \prod_{a=1}^Mq(\boldsymbol{v_a})
\end{equation*}
\\
The Kullback-Leiber divergence between the variational posterior and the true posterior is: 

\begin{align*}
&\text{KL}[q(\boldsymbol{U},\boldsymbol{V},\alpha_0,\alpha_1| \boldsymbol{Y_I}, \boldsymbol{Y_{IA}})|f(\boldsymbol{U},\boldsymbol{V},\alpha_0,\alpha_1 |\boldsymbol{Y_I}, \boldsymbol{Y_{IA}})]\\
&=\int q(\boldsymbol{U},\boldsymbol{V},\alpha_0,\alpha_1 |\boldsymbol{Y_I}, \boldsymbol{Y_{IA}})\log  \frac{q(\boldsymbol{U},\boldsymbol{V},\alpha_0,\alpha_1 |\boldsymbol{Y_I}, \boldsymbol{Y_{IA}})}{f(\boldsymbol{U},\boldsymbol{V},\alpha_0,\alpha_1 |\boldsymbol{Y_I}, \boldsymbol{Y_{IA}})}d(\boldsymbol{U},\boldsymbol{V},\alpha_0,\alpha_1)\\
&=\int \prod_{i=1}^Nq(\boldsymbol{u_i}) \prod_{a=1}^Mq(\boldsymbol{v_a}) 
\log \frac{  \prod_{i=1}^Nq(\boldsymbol{u_i})  \prod_{a=1}^Mq(\boldsymbol{v_a})}{f(\boldsymbol{Y_I}, \boldsymbol{Y_{IA}} |\boldsymbol{U},\boldsymbol{V},\alpha_0,\alpha_1) 
\prod_{i=1}^N f(\boldsymbol{u_i})\prod_{a=1}^M f(\boldsymbol{v_a})}d(\boldsymbol{U},\boldsymbol{V},\alpha_0,\alpha_1)\\
&= \sum_{i=1}^N  \int q(\boldsymbol{u_i}) \log \frac{q(\boldsymbol{u_i})}{f(\boldsymbol{u_i})} d \boldsymbol{u_i} +
\sum_{a=1}^M \int  q(\boldsymbol{v_a}) \log \frac{q(\boldsymbol{v_a})}{f(\boldsymbol{v_a})} d \boldsymbol{v_a}
\\
&-\int q(\boldsymbol{U},\boldsymbol{V},\alpha_0,\alpha_1 |\boldsymbol{Y_I}, \boldsymbol{Y_{IA}})\log f(\boldsymbol{Y_I},\boldsymbol{Y_{IA}} |\boldsymbol{U},\boldsymbol{V},\alpha_0,\alpha_1)d(\boldsymbol{U},\boldsymbol{V},\alpha_0,\alpha_1)\\
&= \sum_{i=1}^N\text{KL}[q(\boldsymbol{u_i})| f(\boldsymbol{u_i})]
+\sum_{a=1}^M\text{KL}[q(\boldsymbol{v_a})| f(\boldsymbol{v_a})]\\
& -E_{q(\boldsymbol{U},\boldsymbol{V},\alpha_0,\alpha_1 |\boldsymbol{Y_I},\boldsymbol{Y_{IA}})}[\log f(\boldsymbol{Y_I}, \boldsymbol{Y_{IA}} |\boldsymbol{U},\boldsymbol{V},\alpha_0,\alpha_1)],
\end{align*}

where each of the components are calculated as follows: 
\begin{align*}
&\sum_{i=1}^N\text{KL}[q(\boldsymbol{u_i})|| f(\boldsymbol{u_i})] &\\
&= -\sum_{i=1}^N \int q(\boldsymbol{u_i}) \log  \frac{f(\boldsymbol{u_i})}{q{(\boldsymbol{u_i})}} d \boldsymbol{u_i}\\
&=- \sum_{i=1}^N \int q(\boldsymbol{u_i}) \Bigg( \frac{1}{2} \bigg( - D \log (\lambda^2_0)+ \log (\det(\tilde{\Lambda}_{0}))- \frac{1}{\lambda^2_0} \boldsymbol{u_i}^T\boldsymbol{u_i} +(\boldsymbol{u_i}-\tilde{\boldsymbol{u_i}})^T \tilde{\Lambda}^{-1}_{0} (\boldsymbol{u_i}-\tilde{\boldsymbol{u_i}})  \bigg)   \Bigg)\\
&=  \frac{1}{2} \Big( DN \log (\lambda^2_0)- N\log (\det(\tilde{\Lambda}_{0})) \Big) + \sum_{i=1}^N \frac{1}{2} \Bigg( \frac{1}{\lambda^2_0} E_{q(\boldsymbol{u_i})}[\boldsymbol{u_i}^T\boldsymbol{u_i}]- E_{q(\boldsymbol{u_i})}[(\boldsymbol{u_i}-\tilde{\boldsymbol{u_i}})^T \tilde{\Lambda}^{-1}_{0} (\boldsymbol{u_i}-\tilde{\boldsymbol{u_i}})]   \Bigg)\\
&= \frac{1}{2} \Big( DN \log (\lambda^2_0)- N\log (\det(\tilde{\Lambda}_{0})) \Big) + \sum_{i=1}^N \frac{1}{2 \lambda^2_0} \Big( \text{Var} (\boldsymbol{u_i}) +\big(E_{q(\boldsymbol{u_i})}[\boldsymbol{u_i}] \big)^2 \Big) -\frac{1}{2}ND\\
&=\frac{1}{2} \Big( DN \log (\lambda^2_0)- N\log (\det(\tilde{\Lambda}_{0})) \Big) +\frac{N \tr(\tilde{\Lambda}_0)}{2 \lambda^2_0} +\frac{\sum_{i=1}^N \boldsymbol{\tilde{u}_i}^T \boldsymbol{\tilde{u}_i}}{2 \lambda^2_0}-\frac{1}{2}ND\\
&\sum_{a=1}^M\text{KL}[q(\boldsymbol{v_a})|| f(\boldsymbol{v_a})]\\
&=\frac{1}{2} \Big( DM \log (\lambda^2_1)- M\log (\det(\tilde{\Lambda}_{1})) \Big) +\frac{M \tr(\tilde{\Lambda}_1)}{2 \lambda^2_1} +\frac{\sum_{a=1}^M \boldsymbol{\tilde{v}_a}^T\boldsymbol{\tilde{v}_a}}{2 \lambda^2_1}-\frac{1}{2}MD
\end{align*}
$E_{q(\boldsymbol{U},\boldsymbol{V} |\boldsymbol{Y_I}, \boldsymbol{Y_{IA}})}[\log f(\boldsymbol{Y_I}, \boldsymbol{Y_{IA}} |\boldsymbol{U},\boldsymbol{V})]$ can be expanded into $6$ components: \\
\begin{align*}
& E_{q(\boldsymbol{U},\boldsymbol{V} |\boldsymbol{Y_I}, \boldsymbol{Y_{IA}})}[\log f(\boldsymbol{Y_I}, \boldsymbol{Y_{IA}} |\boldsymbol{U},\boldsymbol{V})]\\
&=\sum_{i=1}^N \sum_{a=1}^M y_{ia} E_{q(\boldsymbol{U},\boldsymbol{V} |\boldsymbol{Y_I}, \boldsymbol{Y_{IA}})}[\alpha_1-(\boldsymbol{u_i}-\boldsymbol{v_a})^T(\boldsymbol{u_i}-\boldsymbol{v_a})]\\
&+\sum_{i=1}^N \sum_{j=1, j \neq i}^N y_{ij} E_{q(\boldsymbol{U},\boldsymbol{V} |\boldsymbol{Y_I}, \boldsymbol{Y_{IA}})}[\alpha_0-(\boldsymbol{u_i}-\boldsymbol{u_j})^T(\boldsymbol{u_i}-\boldsymbol{u_j})]\\
&- \sum_{i=1}^N \sum_{a=1}^M E_{q(\boldsymbol{U},\boldsymbol{V} |\boldsymbol{Y_I}, \boldsymbol{Y_{IA}})}[\log (1+ \exp(\alpha_1-(\boldsymbol{u_i}-\boldsymbol{v_a})^T(\boldsymbol{u_i}-\boldsymbol{v_a})))]\\
&- \sum_{i=1}^N \sum_{j=1, j \neq i}^N E_{q(\boldsymbol{U},\boldsymbol{V} |\boldsymbol{Y_I}, \boldsymbol{Y_{IA}})}[\log (1+ \exp(\alpha_0 -(\boldsymbol{u_i}-\boldsymbol{u_j})^T(\boldsymbol{u_i}-\boldsymbol{u_j})))]\\
\end{align*}


First $2$ components of $E_{q(\boldsymbol{U},\boldsymbol{V} |\boldsymbol{Y_I}, \boldsymbol{Y_{IA}})}[\log f(\boldsymbol{Y_I}, \boldsymbol{Y_{IA}} |\boldsymbol{U},\boldsymbol{V})]$ are calculated as follows:

\begin{align*}
&\sum_{i=1}^N \sum_{j=1, j \neq i}^N y_{ij} E_{q(\boldsymbol{U},\boldsymbol{V} |\boldsymbol{Y_I}, \boldsymbol{Y_{IA}})}[\alpha_0-(\boldsymbol{u_i}-\boldsymbol{u_j})(\boldsymbol{u_i}-\boldsymbol{u_j})^T]\\
&=\sum_{i=1}^N \sum_{j=1, j \neq i}^N y_{ij} \int \big(\alpha_0- (\boldsymbol{u_i}-\boldsymbol{u_j})(\boldsymbol{u_i}-\boldsymbol{u_j})^T \big)  q(\boldsymbol{u_i}) q(\boldsymbol{u_j}) d ( \boldsymbol{u_i},\boldsymbol{u_j})\\
&=\sum_{i=1}^N \sum_{j=1, j \neq i}^N y_{ij} \Bigg[ \tilde{\alpha}_0 -\int (\boldsymbol{u_i}-\boldsymbol{u_j})(\boldsymbol{u_i}-\boldsymbol{u_j})^T q(\boldsymbol{u_i}) q(\boldsymbol{u_j}) d(\boldsymbol{u_i},\boldsymbol{u_j})       \Bigg]\\
&= \sum_{i=1}^N \sum_{j=1, j \neq i}^N y_{ij} \Bigg[\tilde{\alpha}_0 - \int \sum_{d=1}^D (u_{id} -u_{jd})^2 q(\boldsymbol{u_i}) q(\boldsymbol{u_j}) d(\boldsymbol{u_i},\boldsymbol{u_j})    \Bigg] \\
&= \sum_{i=1}^N \sum_{j=1, j \neq i}^N y_{ij} \Bigg[ \tilde{\alpha}_0 -\bigg[ \sum_{d=1}^D \big[ \int u_{id}^2 q(u_{id}) d u_{id} + \int  u_{jd}^2 q(u_{jd}) du_{jd} - \int \int 2 u_{id} u_{jd} q(u_{id}) q(u_{jd}) du_{id}, du_{jd} \big]  \bigg]  \Bigg] \\
&=\sum_{i=1}^N \sum_{j=1, j \neq i}^N y_{ij} \Bigg[ \tilde{\alpha}_0 -\bigg[ \sum_{d=1}^D \big[ E[ u_{id}^2]  +  E[  u_{jd}^2]  - 2 E[ u_{id}] E[u_{jd}]  \big]  \bigg]  \Bigg]\\
&=\sum_{i=1}^N \sum_{j=1, j \neq i}^N y_{ij} \Bigg[ \tilde{\alpha}_0 -\bigg[ \sum_{d=1}^D \big[ Var[ u_{id}] + E[ u_{id}]^2  + Var[ u_{jd}]+ E[  u_{jd}]^2  - 2 E[ u_{id}] E[u_{jd}]  \big]  \bigg]  \Bigg]\\
&= \sum_{i=1}^N \sum_{j=1, j \neq i}^N y_{ij}  \Bigg[ \tilde{\alpha}_0-2 \tr(\tilde{\Lambda}_{0} )- (\boldsymbol{\tilde{u}_i}-\boldsymbol{\tilde{u}_j})^T(\boldsymbol{\tilde{u}_i}-\boldsymbol{\tilde{u}_j})   \Bigg]
\end{align*}
\newpage

\begin{align*}
&\sum_{i=1}^N \sum_{a=1}^M y_{ia} E_{q(\boldsymbol{U},\boldsymbol{V} |\boldsymbol{Y_I}, \boldsymbol{Y_{IA}})}[\alpha_1-(\boldsymbol{u_i}-\boldsymbol{v_a})(\boldsymbol{u_i}-\boldsymbol{v_a})^T]\\
&=\sum_{i=1}^N \sum_{a=1}^M y_{ia} \int \big(\alpha_1- (\boldsymbol{u_i}-\boldsymbol{v_a})(\boldsymbol{u_i}-\boldsymbol{v_a})^T \big)  q(\boldsymbol{u_i}) q(\boldsymbol{v_a}) d ( \boldsymbol{u_i},\boldsymbol{v_a})\\
&=\sum_{i=1}^N \sum_{a=1}^M y_{ia} \Bigg[ \tilde{\alpha}_1 -\int (\boldsymbol{u_i}-\boldsymbol{v_a})(\boldsymbol{u_i}-\boldsymbol{v_a})^T q(\boldsymbol{u_i}) q(\boldsymbol{v_a}) d(\boldsymbol{u_i},\boldsymbol{v_a})       \Bigg]\\
&= \sum_{i=1}^N \sum_{a=1}^M y_{ia} \Bigg[\tilde{\alpha}_1 - \int \sum_{d=1}^D (u_{id} -v_{ad})^2 q(\boldsymbol{u_i}) q(\boldsymbol{v_a}) d(\boldsymbol{u_i},\boldsymbol{v_a})    \Bigg] \\
&= \sum_{i=1}^N \sum_{a=1}^M y_{ia} \Bigg[ \tilde{\alpha}_1 -\bigg[ \sum_{d=1}^D \big[ \int u_{id}^2 q(u_{id}) d u_{id} + \int  v_{ad}^2 q(v_{ad}) dv_{ad} - \int \int 2 u_{id} v_{ad} q(u_{id}) q(v_{ad}) du_{id}, dv_{ad} \big]  \bigg]  \Bigg] \\
&=\sum_{i=1}^N \sum_{a=1}^M y_{ia} \Bigg[ \tilde{\alpha}_1 -\bigg[ \sum_{d=1}^D \big[ E[ u_{id}^2]  +  E[  v_{ad}^2]  - 2 E[ u_{id}] E[v_{ad}]  \big]  \bigg]  \Bigg]\\
&=\sum_{i=1}^N \sum_{a=1}^M y_{ia} \Bigg[ \tilde{\alpha}_1 -\bigg[ \sum_{d=1}^D \big[ Var[ u_{id}] + E[ u_{id}]^2  + Var[ v_{ad}]+ E[  v_{ad}]^2  - 2 E[ u_{id}] E[v_{ad}]  \big]  \bigg]  \Bigg]\\
&= \sum_{i=1}^N \sum_{a=1}^M y_{ia}  \Bigg[ \tilde{\alpha}_1- \tr(\tilde{\Lambda}_{0} )- \tr(\tilde{\Lambda}_{1} )-(\boldsymbol{\tilde{u}_i}-\boldsymbol{\tilde{v}_a})^T(\boldsymbol{\tilde{u}_i}-\boldsymbol{\tilde{v}_a})   \Bigg]
\end{align*}
The last $2$ expectations of the log functions can be simplified using Jensen's inequality and $E_{q(\boldsymbol{U},\boldsymbol{V} |\boldsymbol{Y_I}, \boldsymbol{Y_{IA}})}[\log f(\boldsymbol{Y_I}, \boldsymbol{Y_{IA}} |\boldsymbol{U},\boldsymbol{V})]$ is now: 
\begingroup
\begin{flalign*}
&E_{q(\boldsymbol{U},\boldsymbol{V} |\boldsymbol{Y_I}, \boldsymbol{Y_{IA}})}[\log f(\boldsymbol{Y_I}, \boldsymbol{Y_{IA}} |\boldsymbol{U},\boldsymbol{V})]\\
\leq & \sum_{i=1}^N \sum_{a=1}^M y_{ia}  \Bigg[ \tilde{\alpha}_1- \tr(\tilde{\Lambda}_{0} )- \tr(\tilde{\Lambda}_{1} )-(\boldsymbol{\tilde{u}_i}-\boldsymbol{\tilde{v}_a})^T(\boldsymbol{\tilde{u}_i}-\boldsymbol{\tilde{v}_a})   \Bigg] \\
+&\sum_{i=1}^N \sum_{j=1, j \neq i}^N y_{ij}  \Bigg[ \tilde{\alpha}_0-2 \tr(\tilde{\Lambda}_{0} )- (\boldsymbol{\tilde{u}_i}-\boldsymbol{\tilde{u}_j})^T(\boldsymbol{\tilde{u}_i}-\boldsymbol{\tilde{u}_j})   \Bigg]\\
&-\sum_{i=1}^N \sum_{a=1}^M E_{q(\boldsymbol{U},\boldsymbol{V} |\boldsymbol{Y_I}, \boldsymbol{Y_{IA}})}[\log (1+ \exp(\alpha_1-(\boldsymbol{u_i}-\boldsymbol{v_a})^T(\boldsymbol{u_i}-\boldsymbol{v_a})))]\\
&-\sum_{i=1}^N \sum_{j=1, j \neq i}^N 
\log (1+ E_{q(\boldsymbol{U},\boldsymbol{V} |\boldsymbol{Y_I}, \boldsymbol{Y_{IA}})}[\exp(\alpha_0 -(\boldsymbol{u_i}-\boldsymbol{u_j})^T(\boldsymbol{u_i}-\boldsymbol{u_j}))])
\end{flalign*}
\endgroup

Recall $\boldsymbol{u_i},\boldsymbol{u_j}$ are $D \times 1$ column vectors. Define $\textbf{u} = \boldsymbol{\tilde{u}_i} -\boldsymbol{\tilde{u}_j}$. Then we have,
$\boldsymbol{u_i}-\boldsymbol{u_j} \overset{iid}{=} N(\textbf{u}, 2\tilde{\Lambda}_0)$, where $\textbf{u}$ is a $D \times 1$ vector and $\tilde{\Lambda}_0$ is an $D \times D$ positive semidefinite matrix. Further define
$\textbf{Z}=(2\tilde{\Lambda}_0)^{-1/2}(\boldsymbol{u_i}-\boldsymbol{u_j}-(\boldsymbol{\tilde{u}_i} -\boldsymbol{\tilde{u}_j}))$. Then clearly 
$\textbf{Z}$ follows $D$ dimensional multivariate standard normal distribution and its density function is given by $f_Z(z)=\frac{1}{\sqrt{2\pi}}\exp(-\frac{1}{2}\textbf{z}^T \textbf{z})$. Consequently, we have
$\boldsymbol{u_i}-\boldsymbol{u_j}=2\tilde{\Lambda}_0^{1/2} \textbf{Z} +\textbf{u}$. 

Therefore, we can reparameterize
\begin{align*}
& E_{q(\boldsymbol{U},\boldsymbol{V} |\boldsymbol{Y_I}, \boldsymbol{Y_{IA}})}[ \exp( -(\boldsymbol{u_i}-\boldsymbol{u_j})^T(\boldsymbol{u_i}-\boldsymbol{u_j}))]\\
&=E_{q(\boldsymbol{U},\boldsymbol{V} |\boldsymbol{Y_I}, \boldsymbol{Y_{IA}})}\Bigg[\exp \Bigg(- \Big(   \textbf{Z}^T (2 \tilde{\Lambda}_0)^{1/2} +\textbf{u}^T
\Big)   \Big(  (2 \tilde{\Lambda}_0)^{1/2}\textbf{Z} +\textbf{u}
\Big)   \Bigg)\Bigg]&\\
&=E_{q(\boldsymbol{U},\boldsymbol{V} |\boldsymbol{Y_I}, \boldsymbol{Y_{IA}})}\Bigg[\exp \Bigg(- \textbf{Z}^T(2\tilde{\Lambda}_0 )\textbf{Z}  -2\textbf{Z}^T (2 \tilde{\Lambda}_0)^{1/2}\textbf{u} - \textbf{u}^T\textbf{u}
)\Bigg)\Bigg]&\\
&= \frac{1}{\sqrt{2 \pi}} \int \exp \Bigg(- \textbf{Z}^T \big(2 \tilde{\Lambda}_0+ \frac{1}{2} \textbf{I}\big) \textbf{Z}  -2\textbf{Z}^T (2 \tilde{\Lambda}_0)^{1/2}\textbf{u} - \textbf{u}^T\textbf{u}
\Bigg) d \textbf{Z}
\end{align*}
Now define $Q= \textbf{u}(2\tilde{\Lambda}_0 +\frac{1}{2} \textbf{I})^{-1}(2 \tilde{\Lambda}_0)^{1/2}$. Then the above integral becomes
\begin{align*}
&\frac{1}{\sqrt{2 \pi}} \int \exp \Bigg(- (\textbf{Z}-Q)^T(2\tilde{\Lambda}_0 +\frac{1}{2} \textbf{I} ) (\textbf{Z}-Q) -\textbf{u}^T\textbf{u} + \textbf{u}^T(2\tilde{\Lambda}_0 +\frac{1}{2} \textbf{I} )^{-1}(2 \tilde{\Lambda}_0)\textbf{u}\Bigg)  d \textbf{Z} \\
&= \exp \Big( -\textbf{u}^T\textbf{u} + \textbf{u}^T(2\tilde{\Lambda}_0 +\frac{1}{2} \textbf{I} )^{-1}(2 \tilde{\Lambda}_0)\textbf{u}\Big) \det(\textbf{I} +4 \tilde{\Lambda}_0)^{-\frac{1}{2}}\\
&= \exp \Big( -\textbf{u}^T(\textbf{I}-(2\tilde{\Lambda}_0 +\frac{1}{2} \textbf{I} )^{-1}(2 \tilde{\Lambda}_0))\textbf{u}\Big) \det(\textbf{I} +4 \tilde{\Lambda}_0)^{-\frac{1}{2}}\\
&= \exp \Big( -\textbf{u}^T (4 \tilde{\Lambda}_0 +\textbf{I})^{-1} \textbf{u}  \Big) \det(\textbf{I} +4 \tilde{\Lambda}_0)^{-\frac{1}{2}} \text{        }.
\end{align*}
The last line follows since for any two invertible matrices $A$ and $B$, if $A+B$ is also invertible, then by \cite{henderson1981deriving}
\[
(A+B)^{-1}=A^{-1} - A^{-1}B(I + A^{-1}B)^{-1}A^{-1}.
\]
Letting $B= 4 \tilde{\Lambda}_0$ and $A=I$ gives:
\begin{equation*}
 E_{q(\boldsymbol{U},\boldsymbol{V} |\boldsymbol{Y_I}, \boldsymbol{Y_{IA}})}[ \exp( -(\boldsymbol{u_i}-\boldsymbol{u_j})^T(\boldsymbol{u_i}-\boldsymbol{u_j}))]= \exp \Big( -(\boldsymbol{\tilde{u}_i}-\boldsymbol{\tilde{u}_j})^T (\textbf{I} + 4\tilde{\Lambda}_{0})^{-1}(\boldsymbol{\tilde{u}_i}-\boldsymbol{\tilde{u}_j})  \Big)   \det(\textbf{I} +4 \tilde{\Lambda}_0)^{-\frac{1}{2}}
\end{equation*}

Recall $\boldsymbol{u_i},\boldsymbol{v_a}$ are $D \times 1$ column vectors. Define $\textbf{u} = \boldsymbol{\tilde{u}_i} -\boldsymbol{\tilde{v}_a}$. Then we have,
$\boldsymbol{u_i}-\boldsymbol{v_a} \overset{iid}{=} N(\textbf{u}, \tilde{\Lambda}_0+\tilde{\Lambda}_1)$, where $\textbf{u}$ is a $D \times 1$ vector and $\tilde{\Lambda}_0$ is an $D \times D$ positive semidefinite matrix. Further define
$\textbf{Z}=(\tilde{\Lambda}_0+\tilde{\Lambda}_1)^{-1/2}(\boldsymbol{u_i}-\boldsymbol{v_a}-(\boldsymbol{\tilde{u}_i} -\boldsymbol{\tilde{v}_a}))$. Then clearly 
$\textbf{Z}$ follows $D$ dimensional multivariate standard normal distribution and its density function is given by $f_Z(z)=\frac{1}{\sqrt{2\pi}}\exp(-\frac{1}{2}\textbf{z}^T \textbf{z})$. Consequently, we have
$\boldsymbol{u_i}-\boldsymbol{v_a}=(\tilde{\Lambda}_0+\tilde{\Lambda}_1)^{1/2} \textbf{Z} +\textbf{u}$. 

Therefore, we can reparameterize
\begin{align*}
& E_{q(\boldsymbol{U},\boldsymbol{V} |\boldsymbol{Y_I}, \boldsymbol{Y_{IA}})}[ \exp( -(\boldsymbol{u_i}-\boldsymbol{v_a})^T(\boldsymbol{u_i}-\boldsymbol{v_a}))]\\
&=E_{q(\boldsymbol{U},\boldsymbol{V} |\boldsymbol{Y_I}, \boldsymbol{Y_{IA}})}\Bigg[\exp \Bigg(- \Big(   \textbf{Z}^T (\tilde{\Lambda}_0+\tilde{\Lambda}_1)^{1/2} +\textbf{u}^T
\Big)   \Big(  (\tilde{\Lambda}_0+\tilde{\Lambda}_1)^{1/2}\textbf{Z} +\textbf{u}
\Big)   \Bigg)\Bigg]&\\
&=E_{q(\boldsymbol{U},\boldsymbol{V} |\boldsymbol{Y_I}, \boldsymbol{Y_{IA}})}\Bigg[\exp \Bigg(- \textbf{Z}^T(\tilde{\Lambda}_0+\tilde{\Lambda}_1 )\textbf{Z}  -2\textbf{Z}^T (\tilde{\Lambda}_0+\tilde{\Lambda}_1)^{1/2}\textbf{u} - \textbf{u}^T\textbf{u}
)\Bigg)\Bigg]&\\
&= \frac{1}{\sqrt{2 \pi}} \int \exp \Bigg(- \textbf{Z}^T \big(\tilde{\Lambda}_0+\tilde{\Lambda}_1+ \frac{1}{2} \textbf{I}\big) \textbf{Z}  -2\textbf{Z}^T (\tilde{\Lambda}_0+\tilde{\Lambda}_1)^{1/2}\textbf{u} - \textbf{u}^T\textbf{u}
\Bigg) d \textbf{Z}
\end{align*}
Now define $Q= \textbf{u}(\tilde{\Lambda}_0+\tilde{\Lambda}_1 +\frac{1}{2} \textbf{I})^{-1}(\tilde{\Lambda}_0+\tilde{\Lambda}_1)^{1/2}$. Then the above integral becomes
\begin{align*}
&\frac{1}{\sqrt{2 \pi}} \int \exp \Bigg(- (\textbf{Z}-Q)^T(\tilde{\Lambda}_0+\tilde{\Lambda}_1 +\frac{1}{2} \textbf{I} ) (\textbf{Z}-Q) -\textbf{u}^T\textbf{u} + \textbf{u}^T(\tilde{\Lambda}_0+\tilde{\Lambda}_1 +\frac{1}{2} \textbf{I} )^{-1}(\tilde{\Lambda}_0+\tilde{\Lambda}_1)\textbf{u}\Bigg)  d \textbf{Z} \\
&= \exp \Big( -\textbf{u}^T\textbf{u} + \textbf{u}^T(\tilde{\Lambda}_0+\tilde{\Lambda}_1 +\frac{1}{2} \textbf{I} )^{-1}(\tilde{\Lambda}_0+\tilde{\Lambda}_1)\textbf{u}\Big) \det(\textbf{I} +2\tilde{\Lambda}_0+2\tilde{\Lambda}_1)^{-\frac{1}{2}}\\
&= \exp \Big( -\textbf{u}^T(\textbf{I}-(\tilde{\Lambda}_0+\tilde{\Lambda}_1 +\frac{1}{2} \textbf{I} )^{-1}(\tilde{\Lambda}_0+\tilde{\Lambda}_1))\textbf{u}\Big) \det(\textbf{I} +2\tilde{\Lambda}_0+2\tilde{\Lambda}_1)^{-\frac{1}{2}}\\
&= \exp \Big( -\textbf{u}^T (\textbf{I} +2\tilde{\Lambda}_0+2\tilde{\Lambda}_1)^{-1} \textbf{u}  \Big) \det(\textbf{I} +2\tilde{\Lambda}_0+2\tilde{\Lambda}_1)^{-\frac{1}{2}} \text{        }.
\end{align*}
The last line follows since for any two invertible matrices $A$ and $B$, if $A+B$ is also invertible, then by \cite{henderson1981deriving}
\[
(A+B)^{-1}=A^{-1} - A^{-1}B(I + A^{-1}B)^{-1}A^{-1}.
\]
Letting $A= I$ and $B=2 \tilde{\Lambda}_0 + 2 \tilde{\Lambda}_1$ gives:
\begin{equation*}
 E_{q(\boldsymbol{U},\boldsymbol{V} |\boldsymbol{Y_I}, \boldsymbol{Y_{IA}})}[ \exp( -(\boldsymbol{u_i}-\boldsymbol{v_a})^T(\boldsymbol{u_i}-\boldsymbol{v_a}))]= \exp \Big( -(\boldsymbol{\tilde{u}_i}-\boldsymbol{\tilde{v}_a})^T (\textbf{I} +2\tilde{\Lambda}_0+2\tilde{\Lambda}_1)^{-1}(\boldsymbol{\tilde{u}_i}-\boldsymbol{\tilde{v}_a})  \Big)   \det(\textbf{I} +2\tilde{\Lambda}_0+2\tilde{\Lambda}_1)^{-\frac{1}{2}}
\end{equation*}

Finally, the Kullback-Leiber divergence between the variational posterior and the true posterior is
\begin{align*}
&\text{KL}[q(\boldsymbol{U},\boldsymbol{V} |\boldsymbol{Y_I}, \boldsymbol{Y_{IA}})||f(\boldsymbol{U},\boldsymbol{V} |\boldsymbol{Y_I}, \boldsymbol{Y_{IA}})]\\
\geq  
& \frac{1}{2} \Big( DN \log (\lambda^2_0)- N\log (\det(\tilde{\Lambda}_{0})) \Big) +\frac{N \tr(\tilde{\Lambda}_0)}{2 \lambda^2_0} +\frac{\sum_{i=1}^N \boldsymbol{\tilde{u}_i}^T\boldsymbol{\tilde{u}_i}}{{2 \lambda^2_0}}-\frac{1}{2}ND\\
+&\frac{1}{2} \Big( DM \log (\lambda^2_1)- M\log (\det(\tilde{\Lambda}_{1})) \Big) +\frac{M \tr(\tilde{\Lambda}_1)}{2 \lambda^2_1} +\frac{\sum_{a=1}^M \boldsymbol{\tilde{v}_a}^T\boldsymbol{\tilde{v}_a}}{2 \lambda^2_1}-\frac{1}{2}MD\\
-& \sum_{i=1}^N \sum_{a=1}^M y_{ia}  \Bigg[ \tilde{\alpha}_1- \tr(\tilde{\Lambda}_{0} )- \tr(\tilde{\Lambda}_{1} )-(\boldsymbol{\tilde{u}_i}-\boldsymbol{\tilde{v}_a})^T(\boldsymbol{\tilde{u}_i}-\boldsymbol{\tilde{v}_a})   \Bigg] \\
-&\sum_{i=1}^N \sum_{j=1, j \neq i}^N y_{ij}  \Bigg[ \tilde{\alpha}_0-2 \tr(\tilde{\Lambda}_{0} )- (\boldsymbol{\tilde{u}_i}-\boldsymbol{\tilde{u}_j})^T(\boldsymbol{\tilde{u}_i}-\boldsymbol{\tilde{u}_j})   \Bigg]\\
+&\sum_{i=1}^N \sum_{a=1}^M   \log \Bigg( 1+\frac{\exp(\tilde{\alpha}_1)}{ \det(\textbf{I} +2\tilde{\Lambda}_0+2\tilde{\Lambda}_1)^{\frac{1}{2}}}\exp \Big( -(\boldsymbol{\tilde{u}_i}-\boldsymbol{\tilde{v}_a})^T (\textbf{I} +2\tilde{\Lambda}_0+2\tilde{\Lambda}_1)^{-1}(\boldsymbol{\tilde{u}_i}-\boldsymbol{\tilde{v}_a})  \Big)  
\Bigg)\\
+&\sum_{i=1}^N \sum_{j=1, j \neq i}^N \log \Bigg( 1+ \frac{\exp(\tilde{\alpha}_0)}{\det (\textbf{I} + 4\tilde{\Lambda}_{0})^{1/2}} \exp \Big( -(\boldsymbol{\tilde{u}_i}-\boldsymbol{\tilde{u}_j})^T (\textbf{I} + 4\tilde{\Lambda}_{0})^{-1}(\boldsymbol{\tilde{u}_i}-\boldsymbol{\tilde{u}_j})  \Big)  \Bigg)+ \text{Const} \boldsymbol{\tilde{u}_i}
\end{align*}

\subsection{Derivations of EM algorithms}
\textbf{E-step}: Estimate $\boldsymbol{\tilde{u}_i}$, $\boldsymbol{\tilde{v}_a}$, $\tilde{\Lambda}_0$ and $\tilde{\Lambda}_1$ by minimizing the KL divergence. 

\begin{align*}
&\text{KL}_{\boldsymbol{\tilde{u}_i}}[q(\boldsymbol{U},\boldsymbol{V} | \boldsymbol{Y_I}, \boldsymbol{Y_{IA}})||f(\boldsymbol{U},\boldsymbol{V} |\boldsymbol{Y_I}, \boldsymbol{Y_{IA}})]\\
\geq  
& \frac{\sum_{i=1}^N \boldsymbol{\tilde{u}_i}^T\boldsymbol{\tilde{u}_i}}{{2 \lambda^2_0}}  \\
-& \sum_{i=1}^N \sum_{a=1}^M y_{ia}  \Bigg[ \tilde{\alpha}_1- \tr(\tilde{\Lambda}_{0} )- \tr(\tilde{\Lambda}_{1} )-(\boldsymbol{\tilde{u}_i}-\boldsymbol{\tilde{v}_a})^T(\boldsymbol{\tilde{u}_i}-\boldsymbol{\tilde{v}_a})   \Bigg] \\
-&\sum_{i=1}^N \sum_{j=1, j \neq i}^N y_{ij}  \Bigg[ \tilde{\alpha}_0-2 \tr(\tilde{\Lambda}_{0} )- (\boldsymbol{\tilde{u}_i}-\boldsymbol{\tilde{u}_j})^T(\boldsymbol{\tilde{u}_i}-\boldsymbol{\tilde{u}_j})   \Bigg]\\
+&\sum_{i=1}^N \sum_{a=1}^M   \log \Bigg( 1+\frac{\exp(\tilde{\alpha}_1)}{ \det(\textbf{I} +2\tilde{\Lambda}_0+2\tilde{\Lambda}_1)^{\frac{1}{2}}}\exp \Big( -(\boldsymbol{\tilde{u}_i}-\boldsymbol{\tilde{v}_a})^T (\textbf{I} +2\tilde{\Lambda}_0+2\tilde{\Lambda}_1)^{-1}(\boldsymbol{\tilde{u}_i}-\boldsymbol{\tilde{v}_a})  \Big)  
\Bigg)\\
+&\sum_{i=1}^N \sum_{j=1, j \neq i}^N \log \Bigg( 1+ \frac{\exp(\tilde{\alpha}_0)}{\det (\textbf{I} + 4\tilde{\Lambda}_{0})^{1/2}} \exp \Big( -(\boldsymbol{\tilde{u}_i}-\boldsymbol{\tilde{u}_j})^T (\textbf{I} + 4\tilde{\Lambda}_{0})^{-1}(\boldsymbol{\tilde{u}_i}-\boldsymbol{\tilde{u}_j})  \Big)  \Bigg)+ \text{Const} \boldsymbol{\tilde{u}_i}
\end{align*}
To find the closed form updates of $\boldsymbol{\tilde{u}_i}$, we use second-order Taylor-expansions of

\begin{align}
\begin{split}\label{fia}
&\boldsymbol{F_{ia}}=\sum_{i=1}^N \sum_{a=1}^M   \log \Bigg( 1+\frac{\exp(\tilde{\alpha}_1)}{ \det(\textbf{I} +2\tilde{\Lambda}_0+2\tilde{\Lambda}_1)^{\frac{1}{2}}}\exp \Big( -(\boldsymbol{\tilde{u}_i}-\boldsymbol{\tilde{v}_a})^T (\textbf{I} +2\tilde{\Lambda}_0+2\tilde{\Lambda}_1)^{-1}(\boldsymbol{\tilde{u}_i}-\boldsymbol{\tilde{v}_a})  \Big)  
\Bigg)\\
\end{split}\\
\begin{split}\label{fa}
&\boldsymbol{F_i}=\sum_{i=1}^N \sum_{j=1, j \neq i}^N \log \Bigg( 1+ \frac{\exp(\tilde{\alpha}_0)}{\det (\textbf{I} + 4\tilde{\Lambda}_{0})^{1/2}} \exp \Big( -(\boldsymbol{\tilde{u}_i}-\boldsymbol{\tilde{u}_j})^T (\textbf{I} + 4\tilde{\Lambda}_{0})^{-1}(\boldsymbol{\tilde{u}_i}-\boldsymbol{\tilde{u}_j})  \Big)  \Bigg)
\end{split}
\end{align}
The gradients of $\boldsymbol{F_i}$ and $\boldsymbol{F_{ia}}$ with respect to $\boldsymbol{\tilde{u}_i}$ are
\begin{align*}
   &\boldsymbol{G_i(\boldsymbol{\tilde{u}_i})} = - 2 (\textbf{I} + 4\tilde{\Lambda}_{0})^{-1} \sum_{j=1,j \neq i}^N (\boldsymbol{\tilde{u}_i}-\boldsymbol{\tilde{u}_j})  \Bigg[ 1+ \frac{\det (\textbf{I} + 4\tilde{\Lambda}_{0})^{1/2}}{\exp (\tilde{\alpha}_0)}\exp \Big( (\boldsymbol{\tilde{u}_i}-\boldsymbol{\tilde{u}_j})^T (\textbf{I} + 4\tilde{\Lambda}_{0})^{-1}(\boldsymbol{\tilde{u}_i}-\boldsymbol{\tilde{u}_j})  \Big)\Bigg]^{-1}\\
   &\boldsymbol{G_{ia}(\boldsymbol{\tilde{u}_i})} = - 2 (\textbf{I} + 2\tilde{\Lambda}_{0}+  2\tilde{\Lambda}_{1})^{-1} \sum_{a=1}^M (\boldsymbol{\tilde{u}_i}-\boldsymbol{\tilde{v}_a}) \\
   & \Bigg[ 1+ \frac{\det (\textbf{I} + \tilde{\Lambda}_{0} + \tilde{\Lambda}_{1})^{1/2}}{\exp (\tilde{\alpha}_1)}\exp \Big( (\boldsymbol{\tilde{u}_i}-\boldsymbol{\tilde{v}_a})^T (\textbf{I} + 2\tilde{\Lambda}_{0} + 2\tilde{\Lambda}_{1} )^{-1}(\boldsymbol{\tilde{u}_i}-\boldsymbol{\tilde{v}_a})  \Big)\Bigg]^{-1}
\end{align*}

The second-order partial derivatives (Hessian matrices) of $\boldsymbol{F_i}, \boldsymbol{F_{ia}}$ with respect to $\boldsymbol{\tilde{u}_i}$ are

\begin{align*}
&\boldsymbol{H_{i}}(\boldsymbol{\tilde{u}_i}) =
 - 2 (\textbf{I} + 4\tilde{\Lambda}_{0})^{-1} \sum_{j=1,j \neq i}^N \left[  1+ \frac{\det (\textbf{I} + 4\tilde{\Lambda}_{0})^{1/2}}{\exp (\tilde{\alpha}_0)}\exp \Big( (\boldsymbol{\tilde{u}_i}-\boldsymbol{\tilde{u}_j})^T (\textbf{I} + 4\tilde{\Lambda}_{0})^{-1}(\boldsymbol{\tilde{u}_i}-\boldsymbol{\tilde{u}_j})  \Big)
\right]^{-1}\\
&\left [\textbf{I} -\frac{ 2 (\boldsymbol{\tilde{u}_i}-\boldsymbol{\tilde{u}_j}) (\boldsymbol{\tilde{u}_i}-\boldsymbol{\tilde{u}_j})^T (\textbf{I} + 4\tilde{\Lambda}_{0})^{-1}
 }{ 1+ \frac{\exp(\tilde{\alpha}_0)}{\det (\textbf{I} + 4\tilde{\Lambda}_{0})^{1/2}} \exp \Big( -(\boldsymbol{\tilde{u}_i}-\boldsymbol{\tilde{u}_j})^T (\textbf{I} + 4\tilde{\Lambda}_{0})^{-1}(\boldsymbol{\tilde{u}_i}-\boldsymbol{\tilde{u}_j})  \Big)}  
\right]\\
&\boldsymbol{H_{ia}}(\boldsymbol{\tilde{u}_i}) =
 - 2 (\textbf{I} +2\tilde{\Lambda}_0+2\tilde{\Lambda}_1)^{-1} \\
 & \sum_{a=1}^M
 \left[  1+ \frac{\det (\textbf{I} +2\tilde{\Lambda}_0+2\tilde{\Lambda}_1)^{1/2}}{\exp (\tilde{\alpha}_1)}
 \exp \Big( (\boldsymbol{\tilde{u}_i}-\boldsymbol{\tilde{v}_a})^T (\textbf{I} +2\tilde{\Lambda}_0+2\tilde{\Lambda}_1)^{-1}(\boldsymbol{\tilde{u}_i}-\boldsymbol{\tilde{v}_a})  \Big)
\right]^{-1}\\
&\left [\textbf{I} -\frac{ 2 (\boldsymbol{\tilde{u}_i}-\boldsymbol{\tilde{v}_a}) (\boldsymbol{\tilde{u}_i}-\boldsymbol{\tilde{v}_a})^T (\textbf{I} +2\tilde{\Lambda}_0+2\tilde{\Lambda}_1)^{-1}
 }{ 1+ \frac{\exp(\tilde{\alpha}_1)}{\det (\textbf{I} +2\tilde{\Lambda}_0+2\tilde{\Lambda}_1)^{1/2}} \exp \Big( -(\boldsymbol{\tilde{u}_i}-\boldsymbol{\tilde{v}_a})^T (\textbf{I} +2\tilde{\Lambda}_0+2\tilde{\Lambda}_1)^{-1}(\boldsymbol{\tilde{u}_i}-\boldsymbol{\tilde{v}_a})  \Big)}  
\right]\\
\end{align*}

\begin{align*}
&KL_{\boldsymbol{\tilde{u}_i}} = \boldsymbol{\tilde{u}_i}^T \big( \frac{1}{2 \lambda_0^2} + \sum_{i=a}^M y_{ia} +  \sum_{j =1, j \neq i}^N  ( y_{ij} +  y_{ji} ) + \boldsymbol{H_i} (\boldsymbol{\tilde{u}_i}) +\frac{1}{2} \boldsymbol{H_{ia}} (\boldsymbol{\tilde{u}_i})
\big) \boldsymbol{\tilde{u}_i} \\
&- 2 \boldsymbol{\tilde{u}_i} \big(\sum_{i=a}^M y_{ia} \boldsymbol{\tilde{v}_a} + \sum_{j =1, j \neq i}^N  ( y_{ij} +  y_{ji} ) \boldsymbol{\tilde{u}_j} - \boldsymbol{G_{i}} (\boldsymbol{\tilde{u}_i})
 - \frac{1}{2}  \boldsymbol{G_{ia}} (\boldsymbol{\tilde{u}_i}) 
 + ( \boldsymbol{H_i} (\boldsymbol{\tilde{u}_i}) + \frac{1}{2} \boldsymbol{H_{ia}} (\boldsymbol{\tilde{u}_i}) ) \boldsymbol{\tilde{u}_i}
  \big). 
\end{align*}

With the Taylor-expansions of the log functions, we can obtain the closed form update rule of $\boldsymbol{\tilde{u}_i}$ by setting the partial derivative of KL equal to $0$. Finally, we have
\begin{align*}
    & \boldsymbol{\tilde{u}_i} =
    \Bigg[ \Bigg( \frac{1}{2 \lambda_0^2} + \sum_{j =1, j \neq i}^N  ( y_{ij} +  y_{ji} )+ \sum_{a =1}^M y_{ia} ) \boldsymbol{I} + \boldsymbol{H_i} (\boldsymbol{\tilde{u}_i}) +  \frac{1}{2}  \boldsymbol{H_{ia}} (\boldsymbol{\tilde{u}_i}) \Bigg]^{-1} \\
    &\Bigg[ \sum_{j =1, j \neq i}^N ( y_{ij} +  y_{ji} )\boldsymbol{\tilde{u}_j} +  \sum_{a =1}^M y_{ia} \boldsymbol{\tilde{v}_a}   - \boldsymbol{G_i} (\boldsymbol{\tilde{u}_i}) +  \Big( \boldsymbol{H_i} (\boldsymbol{\tilde{u}_i}) +  \frac{1}{2} \boldsymbol{H_{ia}} (\boldsymbol{\tilde{u}_i}) \Big) \boldsymbol{\tilde{u}_i}  -  \frac{1}{2} \boldsymbol{G_{ia}} (\boldsymbol{\tilde{u}_i}) \Bigg]
\end{align*}
Similarly, we can obtain the closed form update rule for $\boldsymbol{\tilde{v}_a}$ by taking the second order Taylor-expansion of $\boldsymbol{F_{ia}}$ (see Equation \ref{fia})
The gradient  and Hessian matrix of $ \boldsymbol{F_{ia}}$ with respect to $\boldsymbol{\tilde{v}_a}$ are

\begin{align*}
   &\boldsymbol{G_{ia}(\boldsymbol{\tilde{v}_a})} = - 2 (\textbf{I} + 2\tilde{\Lambda}_{0}+  2\tilde{\Lambda}_{1})^{-1} \sum_{i=1}^N (\boldsymbol{\tilde{v}_a}-\boldsymbol{\tilde{u}_i}) \\
   & \Bigg[ 1+ \frac{\det (\textbf{I} + \tilde{\Lambda}_{0} + \tilde{\Lambda}_{1})^{1/2}}{\exp (\tilde{\alpha}_1)}\exp \Big( (\boldsymbol{\tilde{u}_i}-\boldsymbol{\tilde{v}_a})^T (\textbf{I} + 2\tilde{\Lambda}_{0} + 2\tilde{\Lambda}_{1} )^{-1}(\boldsymbol{\tilde{u}_i}-\boldsymbol{\tilde{v}_a})  \Big)\Bigg]^{-1}\\
   &\boldsymbol{H_{ia}}(\boldsymbol{\tilde{v}_a}) =
 - 2 (\textbf{I} +2\tilde{\Lambda}_0+2\tilde{\Lambda}_1)^{-1} \\
 & \sum_{i=1}^N
 \left[  1+ \frac{\det (\textbf{I} +2\tilde{\Lambda}_0+2\tilde{\Lambda}_1)^{1/2}}{\exp (\tilde{\alpha}_1)}
 \exp \Big( (\boldsymbol{\tilde{u}_i}-\boldsymbol{\tilde{v}_a})^T (\textbf{I} +2\tilde{\Lambda}_0+2\tilde{\Lambda}_1)^{-1}(\boldsymbol{\tilde{u}_i}-\boldsymbol{\tilde{v}_a})  \Big)
\right]^{-1}\\
&\left [\textbf{I} -\frac{ 2 (\boldsymbol{\tilde{v}_a}-\boldsymbol{\tilde{u}_i}) (\boldsymbol{\tilde{v}_a}-\boldsymbol{\tilde{u}_i})^T (\textbf{I} +2\tilde{\Lambda}_0+2\tilde{\Lambda}_1)^{-1}
 }{ 1+ \frac{\exp(\tilde{\alpha}_1)}{\det (\textbf{I} +2\tilde{\Lambda}_0+2\tilde{\Lambda}_1)^{1/2}} \exp \Big( -(\boldsymbol{\tilde{u}_i}-\boldsymbol{\tilde{v}_a})^T (\textbf{I} +2\tilde{\Lambda}_0+2\tilde{\Lambda}_1)^{-1}(\boldsymbol{\tilde{u}_i}-\boldsymbol{\tilde{v}_a})  \Big)}  
\right]\\
\end{align*}

\begin{align*}
KL_{\boldsymbol{\tilde{v}_a}} = \boldsymbol{\tilde{v}_a}^T \big( \frac{1}{2 \lambda_1^2} + \sum_{i=1}^N\sum_{i=a}^M y_{ia} -\frac{1}{2} \boldsymbol{H_{ia}} (\boldsymbol{\tilde{v}_a})
\big) \boldsymbol{\tilde{v}_a} - 2 \boldsymbol{\tilde{v}_a} \big( \sum_{i=1}^N\sum_{i=a}^M y_{ia} \boldsymbol{\tilde{u}_i} - \frac{1}{2}  \boldsymbol{G_{ia}} (\boldsymbol{\tilde{v}_a})  \big). 
\end{align*}

With the Taylor-expansions of the log functions, we can obtain the closed form update rule of $\boldsymbol{\tilde{v}_a}$ by setting the partial derivative of KL equal to $0$. Then, we have
\begin{align*}
& \boldsymbol{\tilde{v}_a} =
    \Bigg[ \Bigg( \frac{1}{2 \lambda_1^2} + \sum_{i =1}^N y_{ia}  \Bigg) \boldsymbol{I}  -  \frac{1}{2} \boldsymbol{H_{ia}} (\boldsymbol{\tilde{v}_a}) \Bigg]^{-1} \\
    &\Bigg[  \sum_{i=1}^N y_{ia} \boldsymbol{\tilde{u}_i}   -  \frac{1}{2}\boldsymbol{G_{ia}} ( \boldsymbol{\tilde{v}_a}) \Bigg] 
\end{align*} 
To find the closed form updates of $\tilde{\Lambda}_0$ and $\tilde{\Lambda}_1$ we used the first-order Taylor-expansions of $\boldsymbol{F_i} $ and $\boldsymbol{F_{ia}}$. The gradients of $\boldsymbol{F_i}$ and $\boldsymbol{F_{ia}}$ with respect to $\tilde{\Lambda}_0$ are: 
\begin{flalign*}
 \boldsymbol{G_i}(\tilde{\Lambda}_0) = &\sum_{i =1}^N \sum_{j=1, j \neq i}^N \Bigg[ 1+ \frac{\det (\textbf{I} + 4\tilde{\Lambda}_{0})^{1/2}}{\exp (\tilde{\alpha}_0)}\exp \Big( (\boldsymbol{\tilde{u}_i}-\boldsymbol{\tilde{u}_j})^T (\textbf{I} + 4\tilde{\Lambda}_{0})^{-1}(\boldsymbol{\tilde{u}_i}-\boldsymbol{\tilde{u}_j})  \Big)
\Bigg]^{-1}\\
& 4 (\textbf{I} + 4\tilde{\Lambda}_{0})^{-1} \Bigg(
 (\boldsymbol{\tilde{u}_i}-\boldsymbol{\tilde{u}_j})(\boldsymbol{\tilde{u}_i}-\boldsymbol{\tilde{u}_j})^T (\textbf{I} + 4\tilde{\Lambda}_{0})^{-1}  -\frac{1}{2} \textbf{I}\Bigg)\\
 \boldsymbol{G_{ia}}(\tilde{\Lambda}_0) = &\sum_{i =1}^N \sum_{a=1}^M \Bigg[ 1+ \frac{\det (\textbf{I} + 2\tilde{\Lambda}_{0}+  2\tilde{\Lambda}_{1})^{1/2}}{\exp (\tilde{\alpha}_1)}\exp \Big( (\boldsymbol{\tilde{u}_i}-\boldsymbol{\tilde{v}_a})^T (\textbf{I} + 2\tilde{\Lambda}_{0}+  2\tilde{\Lambda}_{1})^{-1}(\boldsymbol{\tilde{u}_i}-\boldsymbol{\tilde{v}_a})  \Big)
\Bigg]^{-1}\\
& 2 (\textbf{I} + 2\tilde{\Lambda}_{0}+  2\tilde{\Lambda}_{1})^{-1} \Bigg(
 (\boldsymbol{\tilde{u}_i}-\boldsymbol{\tilde{v}_a})(\boldsymbol{\tilde{u}_i}-\boldsymbol{\tilde{v}_a})^T (\textbf{I} + 2\tilde{\Lambda}_{0}+  2\tilde{\Lambda}_{1})^{-1}  -\frac{1}{2} \textbf{I} \Bigg)
\end{flalign*}

The gradients of $\boldsymbol{F_a}$ and $\boldsymbol{F_{ia}}$ with respect to $\tilde{\Lambda}_1$ are: 
\begin{flalign*}
 \boldsymbol{G_{ia}}(\tilde{\Lambda}_1) = &\sum_{i =1}^N \sum_{a=1}^M \Bigg[ 1+ \frac{\det (\textbf{I} + 2\tilde{\Lambda}_{0}+  2\tilde{\Lambda}_{1})^{1/2}}{\exp (\tilde{\alpha}_1)}\exp \Big( (\boldsymbol{\tilde{u}_i}-\boldsymbol{\tilde{v}_a})^T (\textbf{I} + 2\tilde{\Lambda}_{0}+  2\tilde{\Lambda}_{1})^{-1}(\boldsymbol{\tilde{u}_i}-\boldsymbol{\tilde{v}_a})  \Big)
\Bigg]^{-1}\\
& 2 (\textbf{I} + 2\tilde{\Lambda}_{0}+  2\tilde{\Lambda}_{1})^{-1} \Bigg(
 (\boldsymbol{\tilde{u}_i}-\boldsymbol{\tilde{v}_a})(\boldsymbol{\tilde{u}_i}-\boldsymbol{\tilde{v}_a})^T (\textbf{I} + 2\tilde{\Lambda}_{0}+  2\tilde{\Lambda}_{1})^{-1}  -\frac{1}{2} \textbf{I} \Bigg)
\end{flalign*}

\begin{flalign*}
     &KL_{{\Lambda}_0} = \tr (\tilde{\Lambda}_{0}) \big(  \frac{N}{2 \lambda_0^2} + \sum_{i =1}^N \sum_{a=1}^M y_{ia} + 2 \sum_{i =1}^N \sum_{j =1}^N y_{ij}
     \big) - \frac{N}{2} \log (\det(\tilde{\Lambda}_{0})) +  \boldsymbol{G_{i}}(\tilde{\Lambda}_0) \tilde{\Lambda}_{0} +  \boldsymbol{G_{ia}}(\tilde{\Lambda}_0) \tilde{\Lambda}_{0}\\
        &KL_{{\Lambda}_1} = \tr (\tilde{\Lambda}_{1}) \big(  \frac{M}{2 \lambda_0^2} + \sum_{i =1}^N \sum_{a=1}^M y_{ia} 
     \big) - \frac{M}{2} \log (\det(\tilde{\Lambda}_{1})) + \boldsymbol{G_{ia}}(\tilde{\Lambda}_1) \tilde{\Lambda}_{1}\\
\end{flalign*}

With the Taylor-expansions of the log functions, we can obtain the closed form update rule of $\tilde{\Lambda}_0$  $\tilde{\Lambda}_1$ by setting the partial derivative of KL equal to $0$. Then, we have

\begin{flalign*}
     &\tilde{\Lambda}_0 = \frac{N}{2}
    \Bigg[ \Bigg( \frac{N}{2} \frac{1}{ \lambda_0^2} + 2\sum_{i =1}^N \sum_{j =1}^N y_{ij} + 
    \sum_{i =1}^N \sum_{a=1}^M y_{ia} \Bigg) \boldsymbol{I} +  \boldsymbol{G_i} (\tilde{\Lambda}_0) +  \boldsymbol{G_{ia}} (\tilde{\Lambda}_0) \Bigg]^{-1}&& \\
    &\tilde{\Lambda}_1 = \frac{M}{2}
    \Bigg[ \Bigg( \frac{M}{2}  \frac{1}{ \lambda_1^2} +  \sum_{i =1}^N \sum_{a=1}^M y_{ia}\Bigg) \boldsymbol{I}  + \boldsymbol{G_{ia}} (\tilde{\Lambda}_1) \Bigg]^{-1} 
\end{flalign*}

\textbf{M-step}: Estimate $\tilde{\alpha}_0$, $\tilde{\alpha}_1$ and $\tilde{\alpha}_2$ by minimizing the KL divergence. To find the closed form updates of $\tilde{\alpha}_0$, $\tilde{\alpha}_1$ and $\tilde{\alpha}_2$, we  used second-order Taylor-expansions of the log functions and set the partial derivatives of KL with respects to $\tilde{\alpha}_0$, $\tilde{\alpha}_1$ and $\tilde{\alpha}_2$ as zeros. Then we have 
\begin{flalign*}
     \tilde{\alpha}_0 =&\frac{\sum^N_{i=1}\sum^N_{j\neq i, j=1}y_{ij}- g_i(\tilde{\alpha}_0) + \tilde{\alpha}_0 h_i(\tilde{\alpha}_0 ) }{h_i(\tilde{\alpha}_0 )}&&\\
    \tilde{\alpha}_1 =&\frac{\sum_{i=1}^N\sum_{a=1}^My_{ia}- g_{ia}(\tilde{\alpha}_2) + \tilde{\alpha}_2 h_{ia}(\tilde{\alpha}_2 ) }{h_{ia}(\tilde{\alpha}_2 )}
\end{flalign*} where
\begin{flalign*}
 g_i(\tilde{\alpha}_0) = &\sum_{i =1}^N \sum_{j=1, j \neq i}^N \Bigg[ 1+ \frac{\det (\textbf{I} + 4\tilde{\Lambda}_{0})^{1/2}}{\exp (\tilde{\alpha}_0)}\exp \Big( (\boldsymbol{\tilde{u}_i}-\boldsymbol{\tilde{u}_j})^T (\textbf{I} + 4\tilde{\Lambda}_{0})^{-1}(\boldsymbol{\tilde{u}_i}-\boldsymbol{\tilde{u}_j})  \Big)
\Bigg]^{-1}&&\\
 h_i(\tilde{\alpha}_0) = &\sum_{i =1}^N \sum_{j=1, j \neq i}^N \Bigg[ 1+ \frac{\det (\textbf{I} + 4\tilde{\Lambda}_{0})^{1/2}}{\exp (\tilde{\alpha}_0)}\exp \Big( (\boldsymbol{\tilde{u}_i}-\boldsymbol{\tilde{u}_j})^T (\textbf{I} + 4\tilde{\Lambda}_{0})^{-1}(\boldsymbol{\tilde{u}_i}-\boldsymbol{\tilde{u}_j})  \Big)
\Bigg]^{-1}&&\\
&\Bigg[ 1+ \frac{\exp (\tilde{\alpha}_0)
}{\det (\textbf{I} + 4\tilde{\Lambda}_{0})^{1/2}
}\exp \Big(- (\boldsymbol{\tilde{u}_i}-\boldsymbol{\tilde{u}_j})^T (\textbf{I} + 4\tilde{\Lambda}_{0})^{-1}(\boldsymbol{\tilde{u}_i}-\boldsymbol{\tilde{u}_j})  \Big)
\Bigg]^{-1}\\
 g_{ia}(\tilde{\alpha}_1) = &\sum_{i =1}^N \sum_{a=1}^M \Bigg[ 1+ \frac{\det (\textbf{I} + 2\tilde{\Lambda}_{0}+  2\tilde{\Lambda}_{1})^{1/2}}{\exp (\tilde{\alpha}_1)}\exp \Big( (\boldsymbol{\tilde{u}_i}-\boldsymbol{\tilde{v}_a})^T (\textbf{I} + 2\tilde{\Lambda}_{0}+  2\tilde{\Lambda}_{1})^{-1}(\boldsymbol{\tilde{u}_i}-\boldsymbol{\tilde{v}_a})  \Big)
\Bigg]^{-1}&&\\
 h_{ia}(\tilde{\alpha}_1) = &\sum_{i =1}^N \sum_{a=1}^M \Bigg[ 1+ \frac{\det (\textbf{I} + 2\tilde{\Lambda}_{0}+  2\tilde{\Lambda}_{1})^{1/2}}{\exp (\tilde{\alpha}_1)}\exp \Big( (\boldsymbol{\tilde{u}_i}-\boldsymbol{\tilde{v}_a})^T (\textbf{I} + 2\tilde{\Lambda}_{0}+  2\tilde{\Lambda}_{1})^{-1}(\boldsymbol{\tilde{u}_i}-\boldsymbol{\tilde{v}_a})  \Big)
\Bigg]^{-1}&&\\
&\Bigg[ 1+ \frac{\exp (\tilde{\alpha}_1)
}{\det (\textbf{I} + 2\tilde{\Lambda}_{0}+  2\tilde{\Lambda}_{1})^{1/2}
}\exp \Big(- (\boldsymbol{\tilde{u}_i}-\boldsymbol{\tilde{v}_a})^T (\textbf{I} + 2\tilde{\Lambda}_{0}+  2\tilde{\Lambda}_{1})^{-1}(\boldsymbol{\tilde{u}_i}-\boldsymbol{\tilde{v}_a})  \Big)
\Bigg]^{-1}
\end{flalign*}

\newpage

\end{document}